\documentclass[aps,prb,reprint,twocolumn,amsmath,amssymb,citeautoscript,longbibliography]{revtex4-2}

\usepackage{graphicx}% Include figure files
\usepackage{dcolumn}% Align table columns on decimal point
\usepackage{bm}% bold math
\usepackage{latexsym,epsfig}
\usepackage{graphicx}
\usepackage{verbatim}
\usepackage{comment}
\usepackage{amsmath}
\usepackage{amssymb}
\usepackage{physics}
\usepackage{stmaryrd}
\usepackage{color}
\usepackage{graphicx}
\usepackage{epstopdf}
\usepackage{grffile}
\usepackage{lipsum}
\usepackage{enumitem}

\usepackage{ulem}

\usepackage[dvipsnames]{xcolor}
\DeclareGraphicsExtensions{.eps}

\newcommand{\beq}{\begin{equation}}
\newcommand{\eeq}{\end{equation}}
\newcommand{\bea}{\begin{eqnarray}}
\newcommand{\eea}{\end{eqnarray}}
\newcommand{\ben}{\begin{eqnarray*}}
\newcommand{\een}{\end{eqnarray*}}
\newcommand{\bfig}{\begin{figure}}
\newcommand{\efig}{\end{figure}}

\usepackage{hyperref}
\hypersetup{
    colorlinks=true,      
    urlcolor=blue,
    citecolor=blue,
    linkcolor=blue
}

\usepackage{booktabs}
\usepackage{array}
\usepackage{lipsum} % For dummy text, optional
\usepackage{array} % Add this to your preamble

\begin{document}
\title{Interplay of Electrode Coupling Engineering, Quasiperiodicity, and Magnetic Flux in
Quantum Transport through a Su-Schrieffer-Heeger Ring.}
\author{Sridhar}\email{a22ph09005@iitbbs.ac.in}\thanks{These authors contributed equally to this work.}
\author{Souvik Roy}\email{souvikroy138@gmail.com} \thanks{These authors contributed equally to this work.}
\author{Malay Bandyopadhyay}\email{malay@iitbbs.ac.in}

% \author{Sridhar$^1$}
% \author{Souvik Roy$^1$}
% \author{Malay Bandyopadhyay$^1$}
\affiliation{School of Basic Sciences, Indian Institute of Technology Bhubaneswar, Odisha, India 752050}

\date{\today}

\begin{abstract}
{
We reveal that engineering electrode-coupling configurations can fundamentally reshape coherent transport phenomena in quasiperiodic quantum systems. Leveraging nonequilibrium Green's function theory, we systematically analyze charge and heat transport, as well as current fluctuations, in a magnetic-flux-threaded quasiperiodic Su-Schrieffer-Heeger ring with both symmetric and asymmetric multi-site reservoir couplings. Contrary to the conventional expectation that optimal transport is achieved near the homogeneous-hopping limit, our results reveal that multi-site lead coupling fundamentally reshapes the transport landscape, extending the regime of enhanced transport deep into the topological phase. Strikingly, asymmetric source-drain coupling induces a disorder-assisted conducting phase wherein quasiperiodic modulation enhances, rather than suppresses, charge and energy transport. Magnetic flux exerts a dual influence: it activates
additional interference-mediated transmission channels that amplify transport, while simultaneously suppressing the disorder-induced re-entrant conducting regime. Furthermore, we uncover a flux-driven migration of the optimal transport window with increasing disorder strength, shifting from the topological regime toward the trivial-hopping regime. This behavior highlights the intricate interplay among quasiperiodicity, dimerization, magnetic-flux-induced quantum interference, and the geometry of the system–reservoir coupling. Collectively, our findings position coupling engineering as a powerful paradigm for the rational control of nonequilibrium transport in quasiperiodic materials, and chart a route toward quantum device configurations in which transport characteristics can be precisely tuned via the interplay of disorder, topology, and quantum interference.}
\end{abstract}

\maketitle

\section{Introduction}
%\section{Introduction}

Quantum transport in low-dimensional systems stands at the forefront of mesoscopic physics, underpinned by both its foundational importance and its transformative potential for nanoscale electronics, thermoelectric energy conversion, and quantum information technologies~\cite{Datta1995,Datta2005,Imry2002,Landauer1970,Buttiker1986}. In phase-coherent conductors, charge transport is dictated not only by the intrinsic band structure but also by quantum interference among multiple propagation paths. As a result, transport characteristics become exquisitely sensitive to lattice geometry, disorder, external perturbations, and the architecture of system--reservoir coupling, often manifesting emergent phenomena absent from classical transport theory~\cite{article,Deng,xv5t-qvcm,PhysRevB.97.174206}. Unraveling how these intertwined factors can be harnessed to control charge and energy flow constitutes a central challenge for the rational design of quantum devices.
Quasiperiodic systems offer a fertile landscape for investigating coherent transport phenomena that transcend the binary classification of crystalline order and random disorder. Unlike periodic lattices, quasiperiodic structures lack translational symmetry; unlike random systems, they are characterized by deterministic spatial correlations. This unique intermediate order engenders unconventional spectral and localization properties, positioning quasiperiodic systems as exemplary platforms for probing quantum interference and localization phenomena~\cite{9yh2-lkjp,5vmn-vsxf,Yildiz2025,Maiti2011,PhysRevLett.109.116404}. Notably, the Aubry--Andr\'e--Harper (AAH) model epitomizes this class, exhibiting a tunable localization transition governed by the strength of quasiperiodic modulation and serving as a canonical framework for studies of quantum transport, topological pumping, and localization physics~\cite{AubryAndre1980,Harper1955,Roati2008,Lahini2009,Kraus2012,PhysRevLett.114.146601}.

The Su--Schrieffer--Heeger (SSH) model, a foundational construct in modern condensed-matter physics, was initially formulated to describe the electronic properties of polyacetylene and has since emerged as a prototypical example of a topological lattice system~\cite{Su1979,Su1980,Meier2016, Ryu2002, Asboth2016,Deng2019,Bhattacharya2025,Mazza2014}. Defined by alternating hopping amplitudes, the SSH model supports distinct dimerized phases and features a transport gap arising from bond-order modulation. Beyond its topological attributes, SSH dimerization serves as a robust mechanism for modulating coherent transport by reshaping both lattice connectivity and the underlying spectral landscape. The interplay between SSH dimerization and quasiperiodic modulation introduces a complex competition among localization, gap formation, and quantum interference, leading to unconventional localization profiles, modified spectral characteristics, and nontrivial thermoelectric responses~\cite{Liu2015,Ganeshan2015,Wang2020}.
Magnetic flux represents another crucial tuning parameter for mesoscopic transport. In ring geometries, the Aharonov--Bohm phase introduced by a magnetic field alters the relative phases accumulated along different electronic trajectories, thereby modifying the quantum interference landscape governing coherent transport~\cite{Aharonov1959,Buttiker1983,Gefen1984,l9gd-k9yw,Bedkihal2025}. This flux-controlled modulation of transmission pathways provides a powerful route for manipulating charge and energy transport in mesoscopic architectures. When combined with quasiperiodicity and SSH dimerization, magnetic flux introduces an additional degree of control, enabling interference-induced spectral reconstruction and enriching the transport phase space~\cite{Roy,Mondal,Roy2023}.

Despite significant advances in the understanding of quasiperiodic and topological transport, the role of electrode-coupling engineering remains comparatively unexplored. Most theoretical studies assume that source and drain electrodes couple to the system through a single lattice site. Although this approximation captures many essential transport features, it overlooks the possibility that the electrode-system architecture itself can serve as an independent tuning parameter. Recent developments in molecular electronics, quantum networks, and engineered mesoscopic systems have enabled the realization of extended coupling involving multiple coupling points to external reservoirs~\cite{Aydin2025,Sitek2013,Dicke1953,Wang2013}. 
In phase-coherent systems, such multi-site coupling geometries can substantially modify interference pathways, redistribute transmission channels, and consequently exert profound influence over transport characteristics.
From a fundamental perspective, electrodes play a role beyond merely injecting and extracting carriers; they establish the boundary conditions for quantum interference and determine how electronic wave functions couple to external reservoirs. Therefore, changing the number and spatial arrangement of coupling sites can produce substantial modifications in transport behavior without altering the intrinsic lattice Hamiltonian. While coupling-induced transport modulation has been explored in selected mesoscopic systems~\cite{l9gd-k9yw,mazza2014thermoelectric,PhysRevB.87.045418,PhysRevB.83.115318,PhysRevB.86.115453,PhysRevB.86.195403,PhysRevB.85.155324}, its interplay with quasiperiodicity, SSH dimerization, and magnetic-flux-driven quantum interference remains largely unexplored.

In this work, we present a comprehensive study of charge and heat transport, together with current fluctuations, in a quasiperiodic SSH ring threaded by magnetic flux and connected to external reservoirs through various multi-site electrode configurations. Employing the nonequilibrium Green's function (NEGF) formalism, we systematically investigate both symmetric and asymmetric coupling geometries and elucidate the evolution of the transport landscape under the combined influence of quasiperiodic modulation and Aharonov--Bohm interference.
Our results demonstrate that multi-site electrode coupling, asymmetric source--drain attachment, and magnetic flux collectively reshape and control the transport phase diagram in unexpected ways. In particular, multi-site coupling extends the transport-optimal regime beyond the homogeneous-hopping limit and deep into the topological hopping regime. Asymmetric coupling gives rise to a disorder-assisted transport regime in which moderate quasiperiodic modulation unexpectedly enhances both charge and heat transport. Moreover, magnetic flux activates interference-mediated transmission channels and suppresses the disorder-induced reentrant transport phase under asymmetric coupling. Furthermore, for asymmetric coupling configurations, increasing the quasiperiodic modulation strength drives the transport-optimal window from the topological dimerized regime toward the trivial-hopping regime, revealing a nontrivial competition among electrode geometry, coupling asymmetry, quasiperiodicity, and flux-induced quantum interference.

Overall, our findings establish electrode-coupling engineering as a powerful and versatile strategy for controlling coherent charge and energy transport in quasiperiodic mesoscopic systems. By demonstrating that the electrode-reservoir configuration can influence transport as profoundly as lattice topology, disorder, and magnetic fields, this work opens new possibilities for the design of quantum-interference-based nanoscale devices.
The remainder of the paper is organized as follows. In Sec.~\ref{secII}, we introduce the quasiperiodic SSH ring model and present the theoretical framework based on the nonequilibrium Green's function formalism used to calculate the transmission probability, charge current, heat current, and current noise. Section~\ref{secIII} discusses the transport characteristics for different system-electrode coupling geometries, emphasizing the critical role of electrode engineering in shaping the transport landscape. Finally, the main conclusions are summarized in Sec.~\ref{secIV}.
\section{ Model and Theoretical Formulation}\label{secII}
\subsection{Quasiperiodic SSH Ring and Tight-Binding Description}
We consider a quasiperiodic Su-Schrieffer-Heeger (SSH) ring connected to two external electrodes that act as charge reservoirs. The complete system is described within the nearest-neighbor tight-binding framework, where the total Hamiltonian can be decomposed as

\begin{equation}
H = H_R + H_S + H_D + H_{SR} + H_{DR},
\end{equation}

Here, $H_R$ represents the Hamiltonian of the isolated ring, $H_S$ and $H_D$ correspond to the source and drain electrodes, respectively, while $H_{SR}$ and $H_{DR}$ account for the coupling between the ring and the electrodes.

The effect of an Aharonov--Bohm flux $\phi$ threading the ring is incorporated through the Peierls substitution, whereby each hopping amplitude acquires a phase factor $e^{i\theta}$, where $\theta = 2\pi\phi/(N\phi_{0})$ and $\phi_{0}=h/e$ is the magnetic flux quantum. The resulting Hamiltonian reads
\begin{equation}
H_R=
\sum_{n=1}^{N}\epsilon_n c_n^\dagger c_n
+
\sum_{n=1}^{N/2}
\left[
t_1 e^{i\theta} c_{2n-1}^\dagger c_{2n}
+
t_2 e^{i\theta} c_{2n}^\dagger c_{2n+1}
+\mathrm{H.c.}
\right],
\label{eq:HR}
\end{equation}

where $c_n^\dagger$ ($c_n$) creates (annihilates) an electron at the $n$th lattice site. The parameters $t_1$ and $t_2$ denote the alternating nearest-neighbor hopping amplitudes, giving rise to the characteristic dimerized structure of the SSH lattice. Since the system forms a closed ring, periodic boundary conditions are imposed through the relation $c_{N+1}\equiv c_1$.

To introduce quasiperiodicity, the site energies are modulated according to the Aubry–André–Harper (AAH) prescription,

\begin{equation}
\epsilon_n
=
W\cos(2\pi b n+\phi_{\rm AAH}),
\end{equation}
The quasiperiodic potential introduces deterministic spatial inhomogeneity without invoking stochastic disorder. As the modulation strength (W) increases, the electronic spectrum becomes progressively fragmented, and localization effects become increasingly pronounced. The resulting competition between quasiperiodicity and SSH dimerization constitutes a key ingredient governing the transport properties of the present system.

Here, (W) denotes the modulation amplitude, $b=(\sqrt{5}-1)/2$ is the irrational modulation frequency associated with the golden mean, and $\phi_{\rm AAH}$ is the phase of the quasiperiodic potential, which is fixed at zero throughout this work. The irrational modulation generates a deterministic yet nonperiodic onsite potential landscape, giving rise to a correlated quasidisordered environment that is fundamentally distinct from conventional uncorrelated random disorder.

The source and drain electrodes are described by semi-infinite one-dimensional chains,

\begin{equation}
H_S=
\sum_m \epsilon_0 a_m^\dagger a_m
+
\sum_m t_0
\left(
a_m^\dagger a_{m+1}
+\mathrm{H.c.}
\right),
\end{equation}

and

\begin{equation}
H_D=
\sum_m \epsilon_0 b_m^\dagger b_m
+
\sum_m t_0
\left(
b_m^\dagger b_{m+1}
+\mathrm{H.c.}
\right),
\end{equation}

where $\epsilon_0$ and $t_0$ represent the on-site energy and nearest-neighbor hopping integral within the electrodes.

\begin{figure}
    \centering
    \includegraphics[scale=0.3]{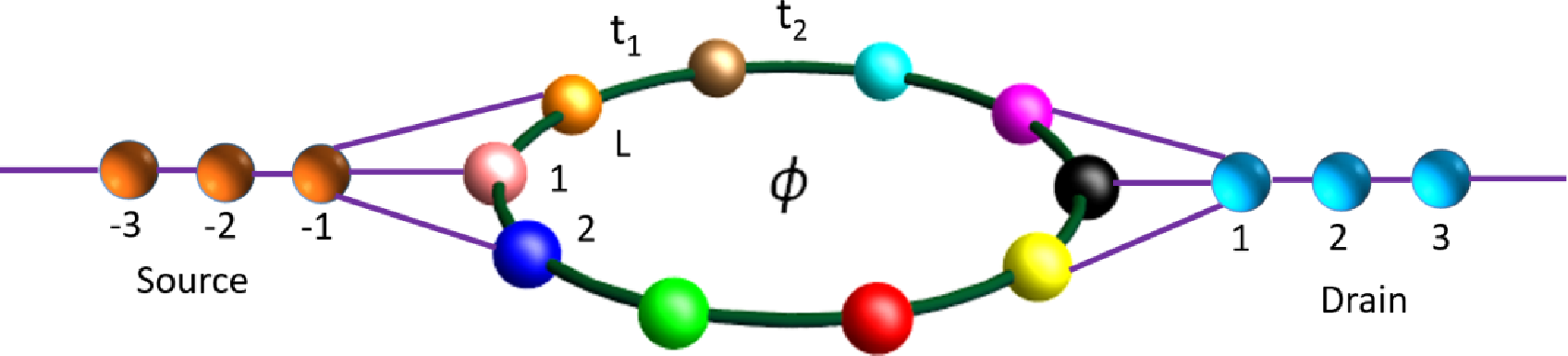}
    \caption{Schematic representation of an SSH ring symmetrically coupled to the source and drain electrodes through three lattice sites at each end. The onsite energies of the ring are subjected to the Aubry–André–Harper (AAH) quasiperiodic modulation.
}
    \label{fig:ring_3_3}
\end{figure}

The coupling between the ring and the electrodes is described by

\begin{equation}
H_{SR}
=
\sum_{p \in \mathcal{S}}
\tau_S
\left(
a_1^\dagger c_p
+\mathrm{H.c.}
\right),
\end{equation}

\begin{equation}
H_{DR}
=
\sum_{q \in \mathcal{D}}
\tau_D
\left(
b_1^\dagger c_q
+\mathrm{H.c.}
\right),
\end{equation}

where $\tau_S$ and $\tau_D$ denote the coupling strengths between the ring and the source and drain electrodes, respectively. Here, $a_1^\dagger$ ($b_1^\dagger$) creates an electron at the terminal site of the source (drain) electrode, while $c_p^\dagger$ and $c_q^\dagger$ correspond to the ring sites coupled to the respective electrodes. The sets $\mathcal{S}$ and $\mathcal{D}$ specify the coupling sites through which the source and drain are attached to the ring. In the symmetric multi-site configuration, three ring sites are connected to the source electrode and three ring sites are connected to the drain electrode (see Fig.~\ref{fig:ring_3_3}). In contrast, the asymmetric coupling geometry is realized by connecting three ring sites to the source while only a single ring site is coupled to the drain (see Fig.~\ref{fig:ring_3_1}). Unlike conventional single-site coupling, multi-site coupling generates off-diagonal hybridization
channels in the effective self-energy matrix. Consequently, electrons injected from the reservoirs can
access multiple interfering pathways simultaneously, allowing the coupling configuration itself to become
an active control parameter for coherent transport.

Having established the microscopic model, we proceed to analyze the transport properties using the nonequilibrium Green's function formalism.
\subsection{Theoretical Framework}
The transport properties of the system are evaluated using the nonequilibrium Green's function (NEGF) technique \cite{dhar2006nonequilibrium}. Within the equation-of-motion formalism, the retarded and advanced Green's functions, $G^-(E)$ and $G^+(E)$, are obtained from the quantum Langevin equations \cite{l9gd-k9yw} describing the dynamics of the system coupled to the external electrodes. The resulting Green's functions are given by
\begin{align}
 G_{i,p(q)}^{\pm}(E)&\nonumber=
\left[
EI-H_R
-\sum_{p^\prime}\Sigma_{p^\prime,p}^{\pm(S)}(E)\delta_{pi}
-\right.\\& \left.\sum_{q^\prime}\Sigma_{q^\prime,q}^{\pm(D)}(E)\delta_{qi}
\right]^{-1}.   
\end{align}
The self-energy matrices encode both the strength and
geometry of the system-reservoir coupling and therefore provide the mathematical origin of the
coupling-engineering effects discussed throughout this work.
Here, $(p,p^\prime)$ and $(q,q^\prime)$ denote the lattice sites coupled to the source ($S$) and drain ($D$) electrodes, respectively. For the symmetric multi-site coupling configuration, where three sites are connected to both the source and drain electrodes, the coupling indices are chosen as $(p,p^\prime)=\{1,2,L\}$ and $(q,q^\prime)=\{L/2 ,L/2+1 ,L/2+2\}$, where $L$ represents the total number of lattice sites in the ring.

In contrast, for the asymmetric multi-site coupling configuration, where the source electrode is connected to three sites while the drain electrode is connected to a single site, the coupling indices become $(p,p^\prime)=\{1,2,L\}$ and $(q,q^\prime)=\{L/2+1\}$. Here, $I$ denotes the identity matrix,  $H_R$ denotes the Hamiltonian of the quantum ring. While $\Sigma_{p^\prime,p}^{\pm(S)}(E)$ and $\Sigma_{q^\prime,q}^{\pm(D)}(E)$ represent the retarded ($+$) and advanced ($-$) self-energy contributions arising from the source and drain electrodes, respectively. Throughout this work we employ the wide-band approximation, in which the electrode density
of states is assumed to vary slowly on the energy scale of the ring spectrum. This approximation
isolates the intrinsic transport properties of the quasiperiodic SSH ring from electrode-specific
spectral features. Within the wide-band limit (WBL), where the density of states of the metallic source and drain electrodes is assumed to be energy independent, the real part of the self-energy vanishes and the self-energy matrices become purely imaginary. Therefore, the source and drain hybridization matrices can be introduced through the relations $\Sigma_{S(D)}^{+}(E)=-\frac{i}{2}\Gamma_{S(D)}$ where $\Gamma_{S(D)}$ characterize the coupling strengths between the ring and the source (drain) electrodes.
The transmission probability of electrons propagating from the source to the drain electrode is expressed as \cite{datta1997electronic}
\begin{equation}
T_{SD}(E,\phi)=\mathrm{Tr}\left[\Gamma_{S}G^{+}(E,\phi)\Gamma_{D}G^{-}(E,\phi)\right],
\end{equation}
where $(\Gamma_{S})$ and $(\Gamma_{D})$ denote the hybridization matrices associated with the source and drain electrodes, respectively, while $(G^{+}(E,\phi))$ and $(G^{-}(E,\phi))$ represent the retarded and advanced Green's functions of the ring system.

We investigate the transport properties of the ring by evaluating the transmission coefficient, charge current, and heat current within the nonequilibrium Green's function formalism. Throughout this work, the source and drain electrodes are taken to have identical coupling strengths, $\gamma_S=\gamma_D=\gamma$, while different source–drain coupling geometries are considered. Throughout this paper, we choose units such that $h = k_B = e = 1$.

The charge transport from the source ($S$) to the drain ($D$) is governed by the transmission probability $T_{SD}(E,\phi)$. Within the Landauer-Büttiker formalism, the charge current is given by
\cite{yamamoto2015thermodynamics,sivan1986multichannel,butcher1990thermal}

Since the transmission probability $T_{SD}(E,\phi)$ constitutes the fundamental transport quantity within the Landauer--B\"uttiker formalism, the average transport properties can be expressed in terms of energy-weighted transmission moments:

\begin{equation}
L_n=\int_{-\infty}^\infty dE\,(E-\mu_S)^n T_{SD}(E,\phi)
\left[f_S(E)-f_D(E)\right].
\label{eq:Ln}
\end{equation}

The zeroth and first moments determine the charge \cite{yamamoto2015thermodynamics,sivan1986multichannel,butcher1990thermal} and heat currents\cite{ benenti2017fundamental,yang2018gate,bergmann2021green,esposito2015nature,esposito2015quantum,mazza2014thermoelectric,seshadri2021entropy,topp2015steady}, respectively,

\begin{equation}
I=L_0,
\qquad
Q_S=L_1.
\end{equation}

Here, $f_{S(D)}(E)$ denotes the Fermi--Dirac distribution function of the source (drain) reservoir, while $\mu_{S(D)}$ and $T_{S(D)}$ represent the corresponding chemical potentials and temperatures. Beyond these average quantities, current fluctuations are characterized by the second cumulant of the transmitted charge distribution. The corresponding zero-frequency current noise is given by \cite{datta1997electronic,PhysRevB.98.155438,PhysRevE.99.062141,PhysRevE.100.042101}

\begin{align}
S &= \int_{-\infty}^{\infty} dE \,
T_{SD}(E,\phi)
\Big\{
f_S(E)\big[1-f_S(E)\big]
\nonumber\\
&\qquad\qquad
+f_D(E)\big[1-f_D(E)\big]
\Big\}
\nonumber\\
&\quad+
\int_{-\infty}^{\infty} dE \,
T_{SD}(E,\phi)
\big[1-T_{SD}(E,\phi)\big]
\nonumber\\
&\qquad\qquad\times
\big[f_S(E)-f_D(E)\big]^2 
\label{eq:noise}
\end{align}

The first term represents thermal (Nyquist) noise, whereas the second term describes shot noise arising from the probabilistic nature of quantum transmission. Thus, charge current, heat current, and current fluctuations all originate from the same transmission function, reflecting different statistical moments and cumulants of the underlying transport process.

While charge and heat currents characterize average transport, current noise provides information about temporal fluctuations and correlations that are inaccessible through mean transport coefficients alone. Noise therefore serves as a sensitive probe of quantum interference and transmission-channel statistics.

\section{Contact Engineered Transport:Results and
Discussions}\label{secIII}

The transport properties of coherent mesoscopic systems are governed not only by their intrinsic electronic structure but also by the manner in which they are coupled to external reservoirs. In this section, we investigate the role of electrode-contact geometry in controlling charge and energy transport through a quasiperiodic SSH ring. By comparing symmetric and asymmetric coupling configurations, we demonstrate that contact architecture acts as an effective tuning parameter that modifies the transmission spectrum, quantum-interference pattern, and energy-filtering characteristics. The interplay among quasiperiodicity, hopping dimerization, and electrode coupling gives rise to distinct transport regimes, revealing new routes for optimizing coherent thermoelectric performance.

In all numerical calculations presented in this work, we fix the system size to $N=34$. The chosen system size provides an adequate rational approximation to the irrational
modulation frequency associated with the golden mean while remaining computationally
efficient. We have verified that the principal transport features remain unchanged for larger
system sizes. The relative strength of the alternating hopping amplitudes is characterized by the parameter $\alpha= t_1/t_2$, The parameter $\alpha$ continuously interpolates between strongly dimerized and nearly homogeneous
hopping configurations and therefore serves as the primary control parameter for characterizing the SSH modulation. The other model parameters are specified in the corresponding figures and discussions.
\subsection{Symmetric Multi-Site Source–Drain Coupling Geometry} 
\subsubsection{Energy resolved transmission spectra}
\begin{figure}[t]
    \centering

    \includegraphics[width=8cm,height=8cm]{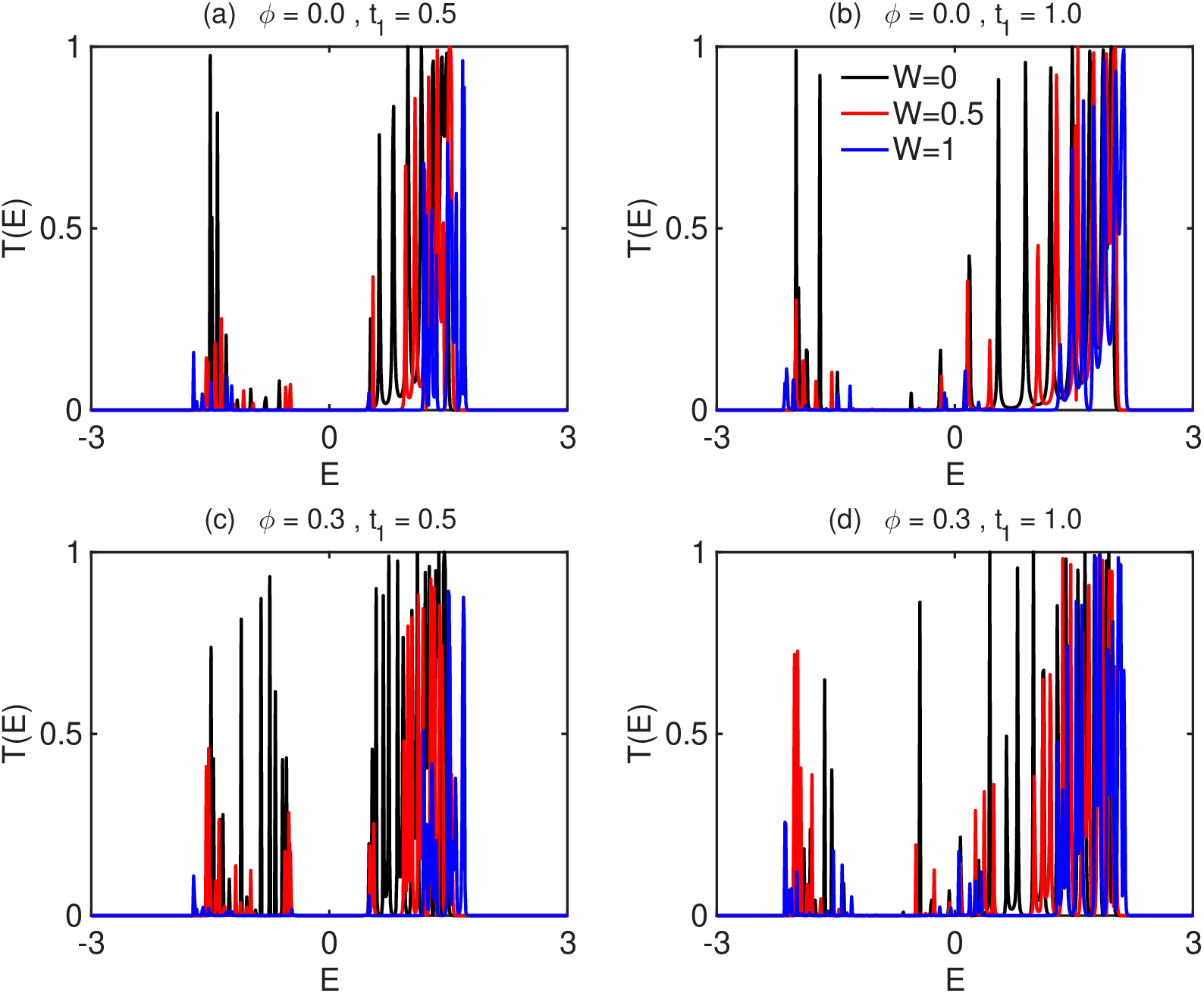}
    \caption{Variation of the transmission probability as a function of energy for the three-site source-three-site drain coupling configuration. The upper and lower rows correspond to the cases with and without magnetic flux, respectively. The first and second columns represent $t_1=0.5$ and $t_1=1.0$, respectively, with $t_2=1$. The black, red, and blue curves correspond to three different values of the quasiperiodic disorder strength $W$, as indicated in the figure. Parameters used: $\gamma=0.05$.}
    \label{fig:1}
\end{figure}
\begin{figure}[t]
    \centering
    \includegraphics[width=8cm,height=10cm]{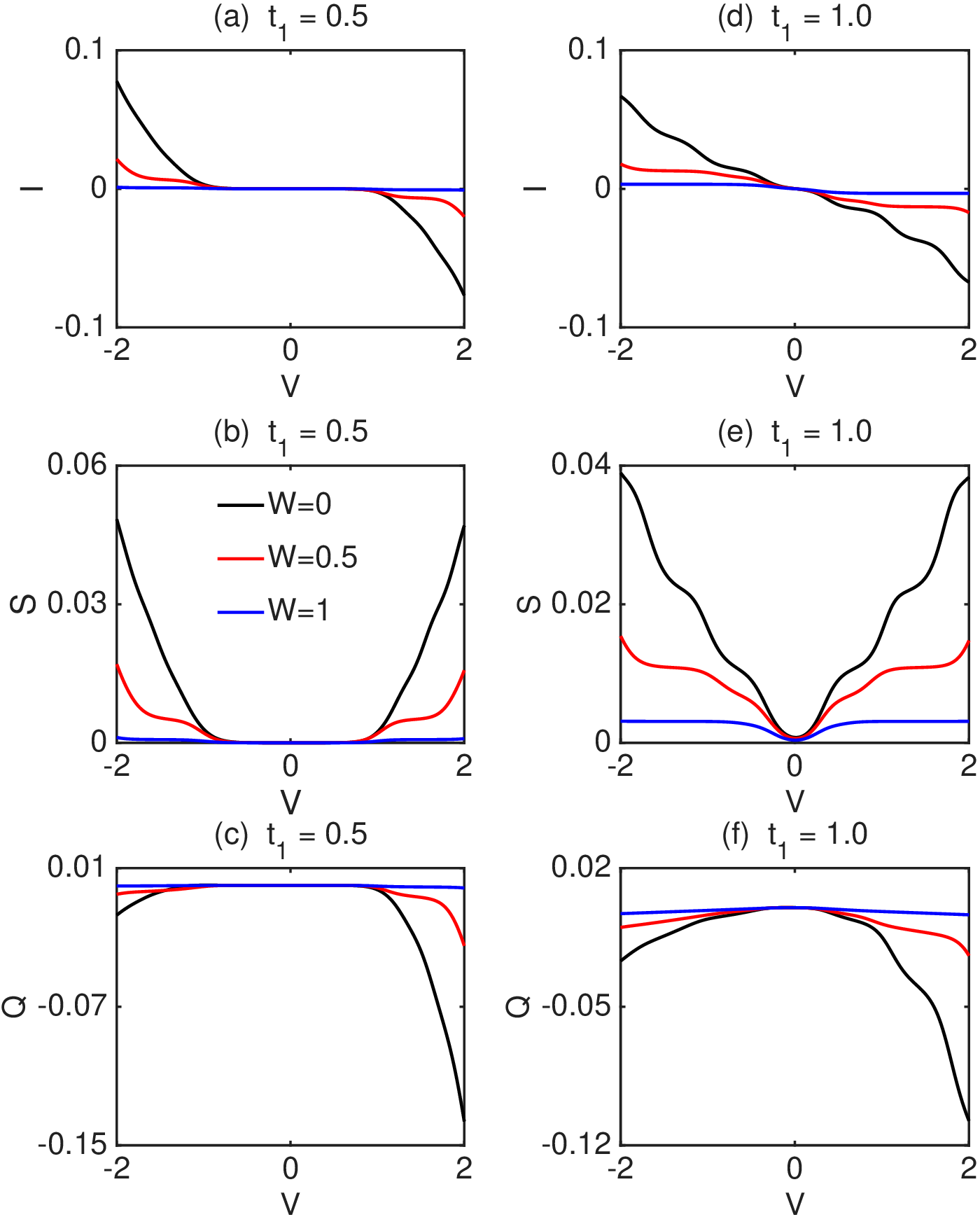}

\caption{
(Color online) Bias dependence Current current [(a),(d)], current noise [(b),(e)], and heat current [(c),(f)] for the multi-site ring configuration in which three lattice sites are coupled to both the source and drain electrodes. The top, middle, and bottom rows correspond to the
charge current, current noise, and heat current, respectively, while the left and right columns correspond to $t_1=0.5$ ($t_1<t_2$) and $t_1=1.0$ ($t_1=t_2$), respectively, with $t_2=1$. Results are shown for different quasiperiodic modulation strengths $W$, as indicated in the figure. Other parameters are $\gamma=0.05 ,\phi=0$, $T_S=T+\Delta T$, $T_D=T-\Delta T$, $T=0.05$, $\Delta T=0.005$, and $\mu_{S,D}=\mp V/2$.
}
\label{fig:2}
\end{figure}

\begin{figure}[t]
    \centering
    \includegraphics[width=8cm,height=10cm]{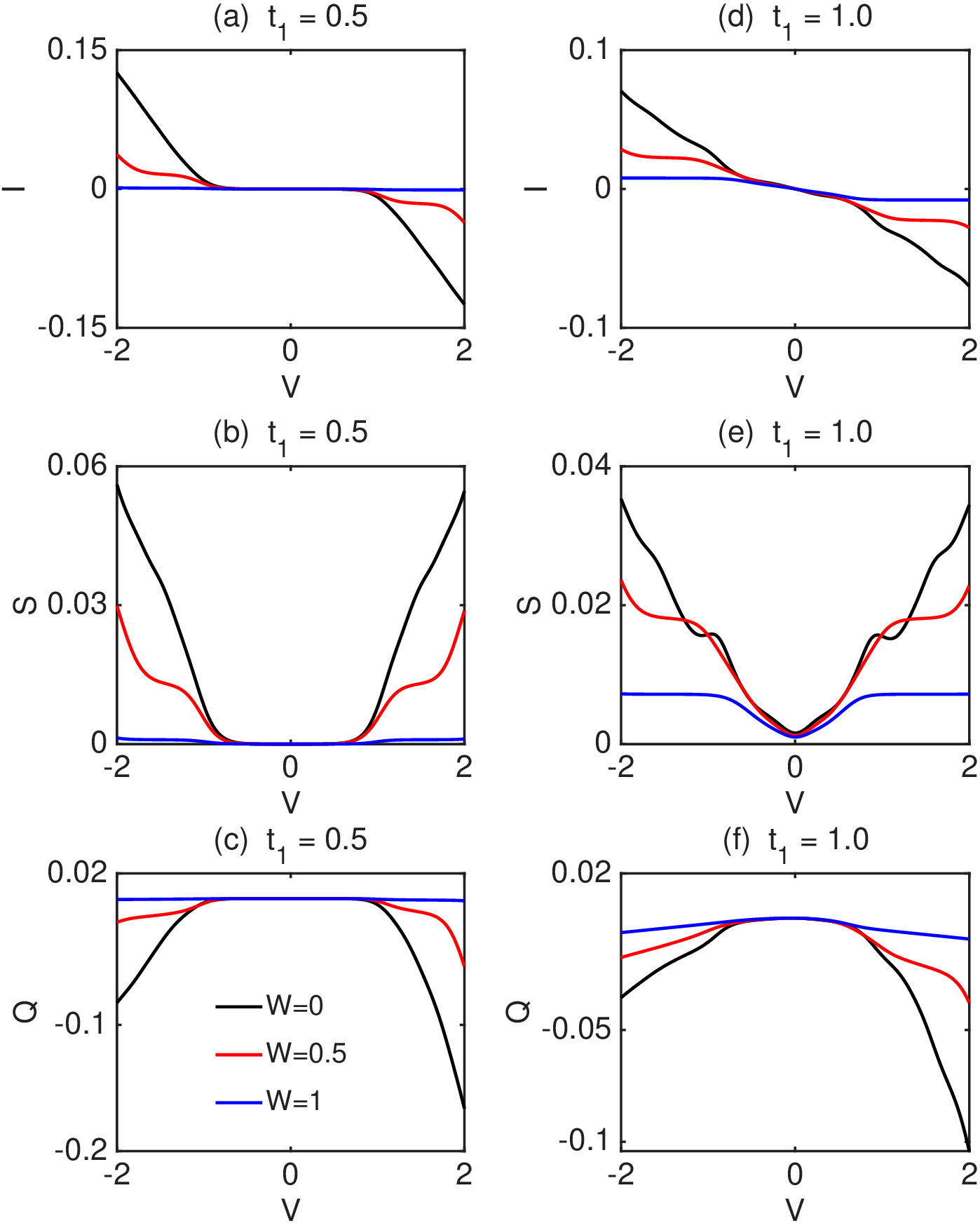}
    
\caption{
(Color online) Bias dependence of the charge current, current noise, and heat current for the symmetric three-site source–three-site drain coupling geometry in the presence of a finite magnetic flux ($\phi=0.3$). All other parameters are identical to those in Fig.~\ref{fig:2}.
}
 \label{fig:3}
\end{figure}

Figure~\ref{fig:1} presents the energy-resolved transmission probability for the symmetric multi-site coupling geometry under different quasiperiodic modulation strengths and magnetic-flux conditions. For the homogeneous hopping configuration ($t_1=t_2$), the transmission spectrum exhibits several resonant features in the vicinity of the Fermi energy, indicating the presence of multiple transport-active states that efficiently couple to the electrodes. As the hopping amplitudes become unequal ($t_1<t_2$), SSH dimerization progressively reconstructs the energy spectrum and opens a transport gap around $E=0$. The resulting suppression of transmission near the band center reflects the reduced availability of conducting states within the dimerization-induced gap. Away from this gap region, the transmission spectrum consists of a sequence of sharp resonances associated with the discrete eigenstates of the finite ring. These resonant channels arise from coherent tunneling through the corresponding energy levels and govern the subsequent charge and heat transport characteristics. Owing to the approximate particle-hole symmetry of the SSH lattice, similar resonant structures also appear at negative energies. However, their spectral weight and density are generally smaller than those observed at positive energies, leading to a comparatively weaker contribution to transport.

The introduction of a magnetic flux profoundly restructures the transmission landscape of the quasiperiodic SSH ring. As evident from Fig.~2, the flux generates a substantially richer resonance spectrum, manifested by the emergence of additional transmission channels across a broad energy window. This enhancement is not confined to the vicinity of the Fermi level but extends throughout both the positive- and negative-energy sectors of the spectrum. Microscopically, the magnetic flux imprints an Aharonov--Bohm phase on the electronic wave function, thereby modifying the relative phases accumulated along different propagation pathways around the ring. The resulting reorganization of quantum-interference conditions reconstructs the spectral coupling between the ring eigenstates and the external reservoirs. Consequently, transport channels that remain weakly transmitting or completely suppressed in the zero-flux configuration become activated, while existing resonances may undergo significant enhancement. The observed increase in the density of transmission resonances therefore reflects a flux-induced, interference-driven spectral reconstruction, which promotes coherent electron propagation and substantially broadens the range of transport-active states.

\subsubsection{Transmission driven transport response and current fluctuations}

% \begin{figure}[t]
%     \centering
%     \includegraphics[width=9cm,height=6cm]{fig/ISQ_source_multi_drain_single_phi_eq_0.eps}
%     \caption{Same as Fig.~2, but for the asymmetric coupling configuration where three lattice sites are connected to the source electrode and a single lattice site is connected to the drain electrode. The magnetic flux is set to $\phi=0$}
%     \label{fig:8}
% \end{figure}

% \begin{figure}[t]
%     \centering
%     \includegraphics[width=9cm,height=6cm]{fig/ISQ_source_multi_drain_single_phi_eq_0.3.eps}
%     \caption{Similar to Fig.~5, but for $\phi=0.3$.}
%     \label{fig:15}
% \end{figure}

Figure~3 presents the bias-voltage dependence of the charge current, current noise, and heat current for the symmetric multi-site coupling geometry. The left and right column correspond to the topological ($t_1<t_2$) and homogeneous ($t_1=t_2$) hopping configurations, respectively, while the three rows display the charge current, current fluctuations, and heat current. For each transport observable, results are shown for three representative quasi-periodic modulation strengths ($W$).

A prominent feature common to all panels is the strong suppression of transport and fluctuation signals in the low-bias regime. This behavior can be directly traced to the transmission characteristics discussed in Fig.~\ref{fig:1}. Near the band center, the transmission probability is either completely absent or significantly reduced due to the combined effects of SSH dimerization, quasi-periodic spectral restructuring, and quantum interference. Since only those states lying within the bias window contribute to transport, the scarcity of transmission-active channels around the Fermi energy results in vanishingly small charge and heat currents, accompanied by a corresponding suppression of current fluctuations. The low-bias transport behavior therefore provides a direct manifestation of the underlying transmission gap and the limited availability of conducting states in the vicinity of the chemical potential.

For the topological configuration ($t_1<t_2$), finite charge and energy transport emerge primarily within the bias range $1\leq V\leq2$, where the bias window encompasses resonant states located outside the dimerization-induced gap. Within this regime, the charge current, heat current, and current fluctuations decrease systematically with increasing quasi-periodic modulation. This suppression arises from the progressive fragmentation of the transmission spectrum, which reduces both the density and spectral weight of transport-active resonances. As a result, fewer electronic states contribute effectively to transport, leading to a simultaneous reduction of the average currents and their associated fluctuations.

A qualitatively different behavior is observed in the homogeneous-hopping regime $t_1=t_2$, where the charge current, current noise, and heat current are found to be somewhat lower than those in the topological dimerized regime, despite the absence of a dimerization gap. This seemingly counterintuitive behavior can be traced to the structure of the transmission spectrum: in the topological regime, the transmission resonances become more densely distributed within the relevant energy window, thereby increasing the number of transport-active channels. The enhancement originates from a dimerization-induced reconstruction of the transmission spectrum, which generates additional resonant states contributing to transport within the bias window. Consequently, electrons gain access to multiple coherent transmission pathways, leading to enhanced charge and energy transport. These findings reveal a disorder-assisted conducting regime in which quasi-periodicity promotes, rather than suppresses, transport through interference-mediated resonant conduction.

Figure~\ref{fig:3} displays the corresponding transport characteristics in the presence of magnetic flux. The flux substantially enhances all transport observables by modifying the interference conditions through the Aharonov--Bohm effect and activating additional resonant transmission channels. In the dimerized phase, transport remains constrained by the SSH gap but is noticeably amplified by the increased density of conducting states. For ($t_1=t_2$), however, the magnetic flux alters the disorder dependence of transport by redistributing the transport-active states and modifying the resonance structure. The resulting behavior demonstrates that quasi-periodicity and magnetic-flux-induced interference cooperate and compete in reshaping the transmission landscape, thereby providing an effective means of controlling coherent charge and energy transport in the system.

\subsection{Asymmetric Multi-Site Source–single site Drain Coupling Geometry}
\begin{figure}[]
    \centering
    \includegraphics[scale=0.3]{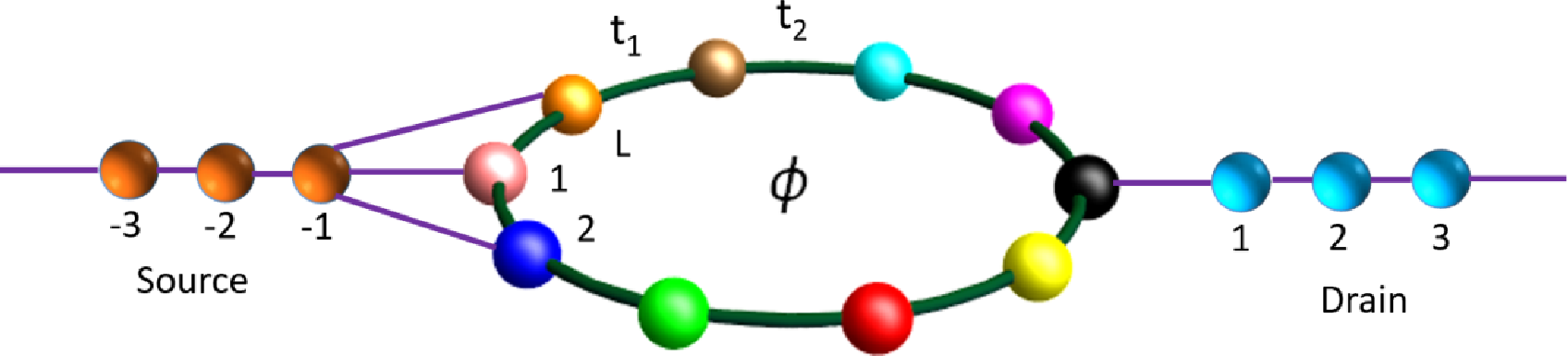}
    \caption{Schematic illustration of an SSH ring with asymmetric source-drain coupling, where three sites are attached to the source electrode and a single site is attached to the drain electrode.
}
    \label{fig:ring_3_1}
\end{figure}

\begin{figure}[]
    \centering
    \includegraphics[width=8cm,height=8cm]{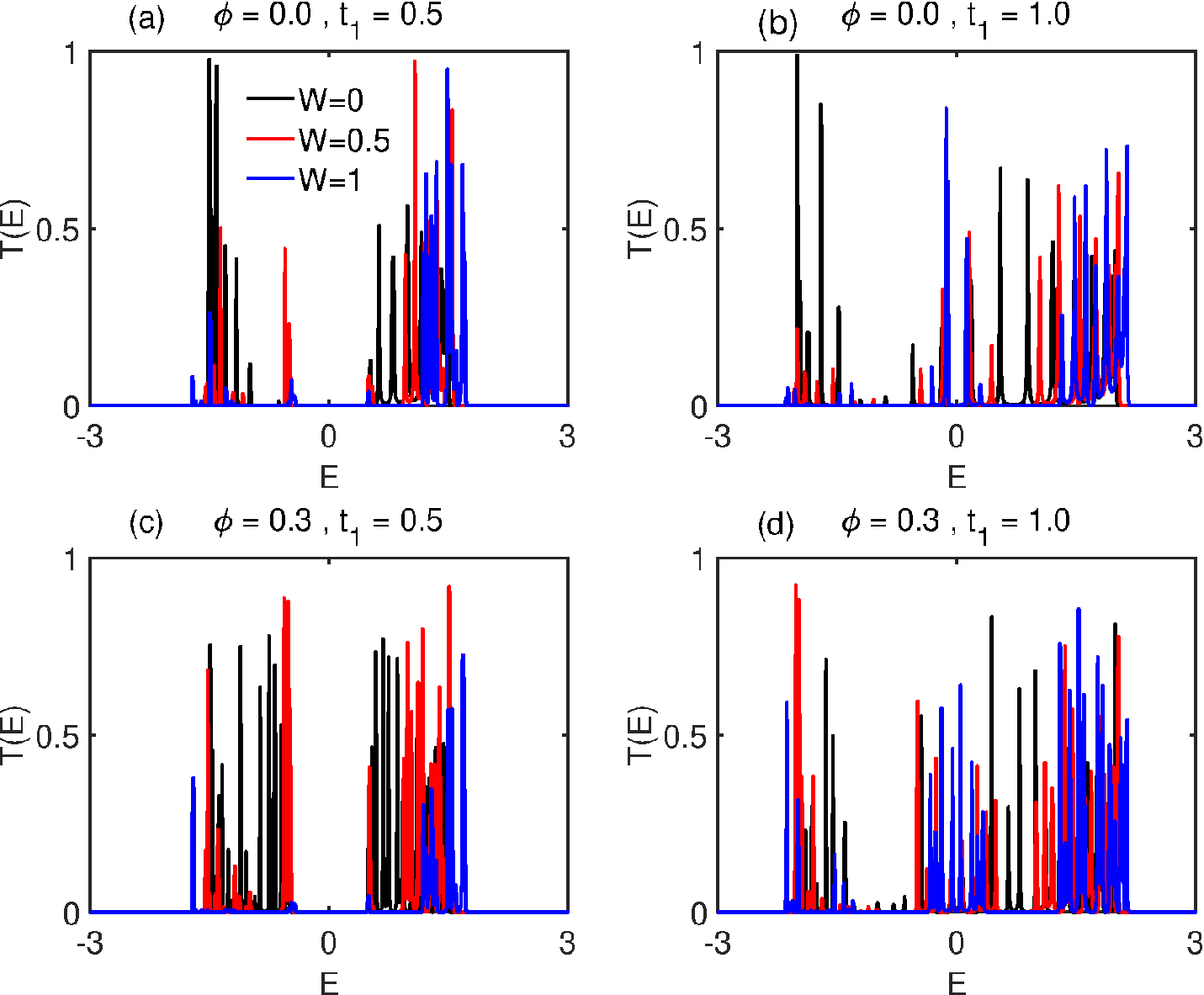}
    
 \caption{Variation of the transmission probability as a function of energy for the asymmetric electrode-coupling configuration, where three sites are connected to the source electrode and a single site is connected to the drain electrode. The upper and lower rows correspond to the cases without and with magnetic flux, respectively. The first and second columns represent $t_1=0.5$ and $t_1=1.0$, respectively, with $t_2=1$. The black, red, and blue curves correspond to different values of the quasi-periodic disorder strength $W$, as indicated in the figure. Parameters used: $\gamma=0.05$.
}

%    \caption{Same as Fig.~\ref{fig:1}, but for the asymmetric coupling configuration where three lattice sites are connected to the source electrode and a single lattice site is connected to the drain electrode.}
    \label{fig:7}
\end{figure}
\begin{figure}[t]
    \centering
    \includegraphics[width=8cm,height=10cm]{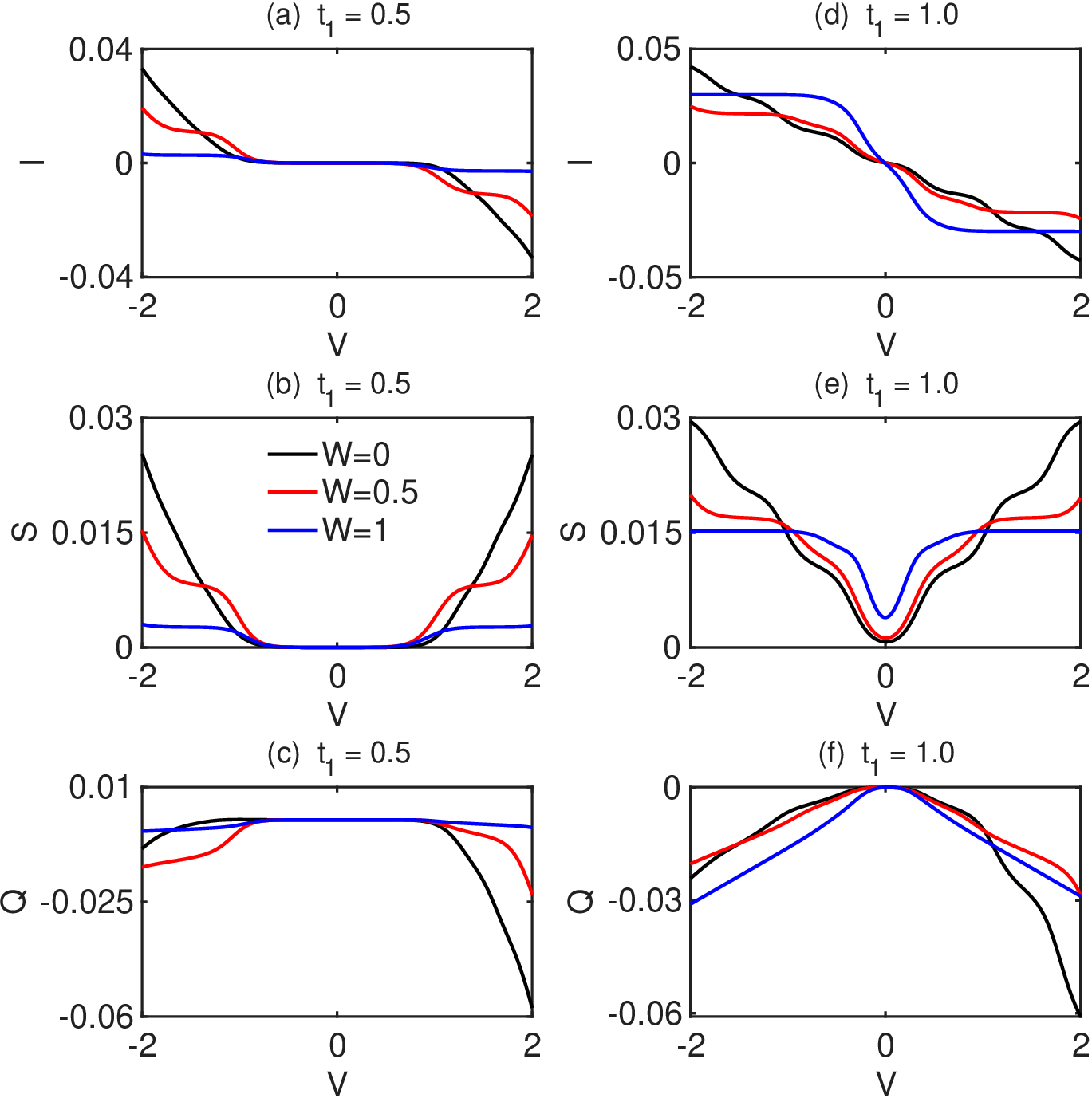}
   
\caption{
(Color online) Bias-voltage dependence of the charge current [(a),(d)], current noise [(b),(e)], and heat current [(c),(f)] for the asymmetric coupling geometry, where three lattice sites are coupled to the source electrode and a single lattice site is coupled to the drain electrode. The first, second, and third rows correspond to the charge current, current noise, and heat current, respectively, while the left and right columns correspond to $t_1=0.5$ ($t_1<t_2$) and $t_1=1.0$ ($t_1=t_2$), respectively, with $t_2=1$. Results are shown for different quasi-periodic modulation strengths $W$, as indicated in the figure. Other parameters are $\gamma=0.05,\phi=0$, $T_S=T+\Delta T$, $T_D=T-\Delta T$, $T=0.05$, $\Delta T=0.005$, and $\mu_{S,D}=\mp V/2$.
}

    % \caption{Same as Fig.~\ref{fig:2}, but for the asymmetric coupling configuration where three lattice sites are connected to the source electrode and a single lattice site is connected to the drain electrode. The magnetic flux is set to $\phi=0$}
    \label{fig:8}
\end{figure}
\begin{figure}[t]
    \centering
    \includegraphics[width=8cm,height=10cm]{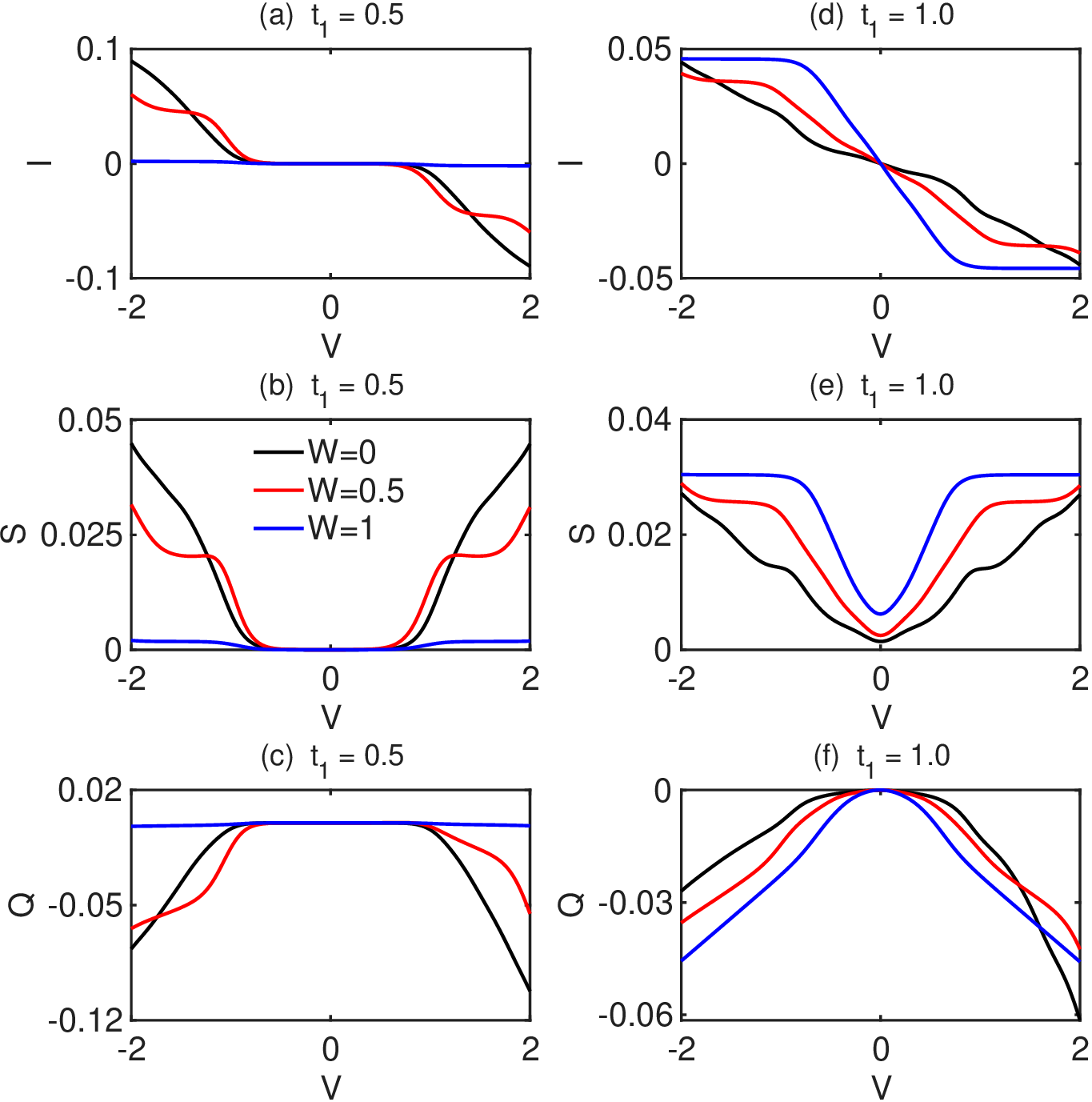}
    \caption{(Color online) Bias-voltage dependence of the charge current [(a),(d)], current noise [(b),(e)], and heat current [(c),(f)] for the asymmetric electrode-coupling configuration, where three lattice sites are connected to the source electrode and a single lattice site is connected to the drain electrode. The left and right columns correspond to $t_1=0.5$ ($t_1<t_2$) and $t_1=1.0$ ($t_1=t_2$), respectively, with $t_2=1$. Results are shown for different quasi-periodic disorder strengths $W$, as indicated in the figure. Other Parameters are identical to those in Fig.~\ref{fig:8}, except $\phi=0.3$.}
    \label{fig:15}
\end{figure}
% \begin{figure}[]
%     \centering
%     \includegraphics[scale=0.3]{fig/Ring_3_1_AAH.eps}
%     \caption{Schematic illustration of an SSH ring with asymmetric source-drain coupling, where three sites are attached to the source electrode and a single site is attached to the drain electrode.
% }
%     \label{fig:ring_3_1}
% \end{figure}
\begin{figure*}[]
    \centering
    \includegraphics[width=15cm,height=4cm]{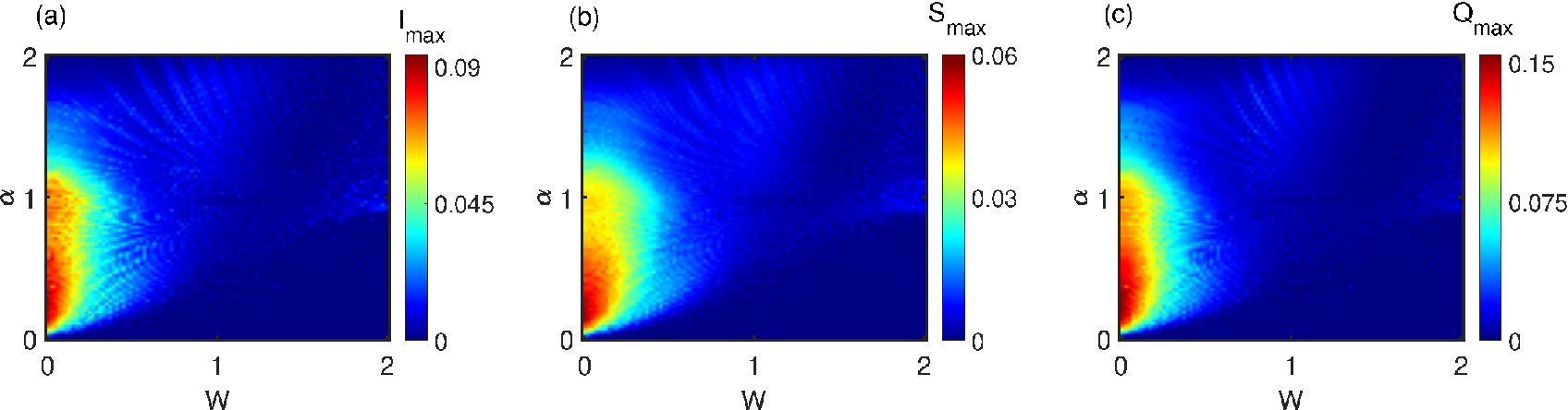}
    
\caption{
(Color online) Color map for symmetric coupling geometry of the maximum (a) charge current, (b) current noise, and (c) heat current in the $\alpha$-$W$ parameter space, obtained by extracting the maximum values within the bias-voltage window $1 \leq V \leq 2$. The magnetic flux is fixed at $\phi=0$. Here, $\alpha=t_1/t_2$ characterizes the degree of SSH dimerization and $W$ denotes the quasi-periodic modulation strength. Other parameters are $\gamma=0.05, T_S=T+\Delta T$, $T_D=T-\Delta T$, $T=0.05$, $\Delta T=0.005$, $t_2=1$, and $\mu_{S,D}=\mp V/2$.
}
\label{fig:4}
\end{figure*}
\begin{figure*}[]
    \centering
    \includegraphics[width=15cm,height=4cm]{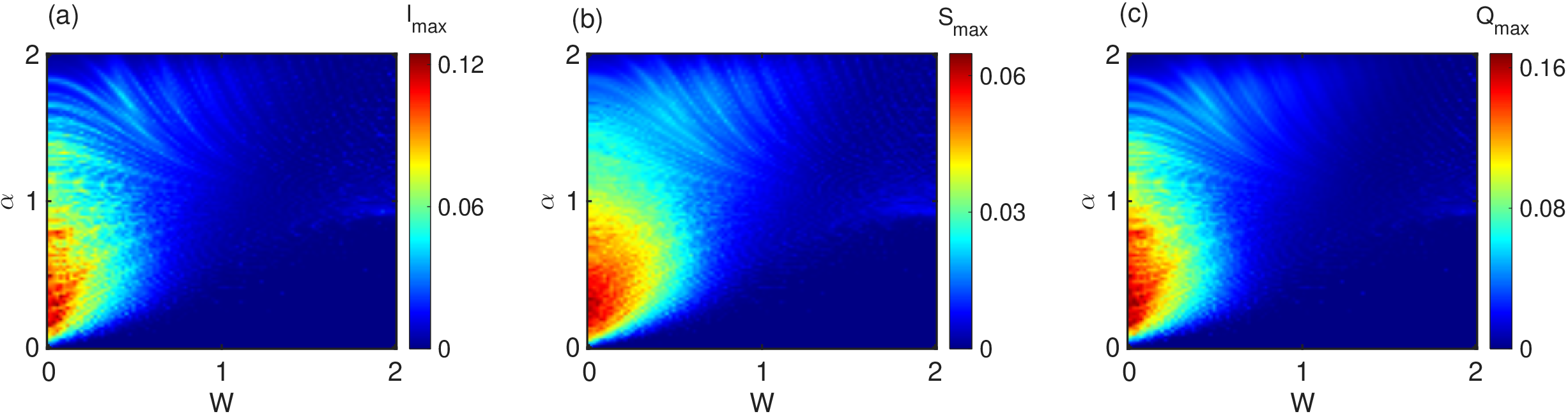}
    \caption{(Color online) Color maps of the maximum (a) charge current, (b) current noise, and (c) heat current in the $\alpha$--$W$ parameter space for the symmetric electrode-coupling configuration, where three lattice sites are connected to both the source and drain electrodes. The maximum values are extracted within the bias-voltage window $1 \leq V \leq 2$. Here, $\alpha=t_1/t_2$ denotes the degree of SSH dimerization and $W$ is the quasi-periodic modulation strength. Other parameters are identical to those in Fig.~\ref{fig:4}, except that $\phi=0.3$.
}
\label{fig:5}
\end{figure*}

\subsubsection{Energy resolved transmission spectra}

Figure~\ref{fig:7} displays the energy-dependent transmission spectrum for the asymmetric coupling configuration, where the source electrode is coupled to three lattice sites while the drain electrode remains connected to a single site. The upper and lower panels correspond to the zero-flux ($\phi=0$) and finite-flux ($\phi=0.3$) configurations, respectively, while the left and right columns represent the dimerized ($t_1<t_2$) and homogeneous ($t_1=t_2$) hopping regimes. Results are shown for several quasiperiodic modulation strengths in order to elucidate the combined influence of coupling asymmetry, hopping modulation, disorder, and magnetic flux on coherent transport.

In the absence of magnetic flux, the dimerized phase [Fig.~\ref{fig:7}(a)] exhibits a pronounced transport gap around the band center that persists over the entire range of quasi-periodic modulation strengths considered. The near-vanishing transmission within this energy window reflects the absence of transport-active states and constitutes a direct manifestation of the SSH dimerization gap. Outside the gap region, transport occurs through a discrete set of resonant channels associated with the eigen-states of the finite ring. Increasing quasi-periodic modulation progressively restructures these resonances, reducing their spectral weight and thereby suppressing coherent transmission. These observations indicate that, in the dimerized regime, quasi-periodicity primarily acts to reinforce the transport-suppressing effects of the underlying SSH gap.

A markedly different behavior emerges in the homogeneous hopping regime [Fig.~\ref{fig:7}(b)], where the removal of the dimerization gap allows conducting states to populate the vicinity of the Fermi energy. Consequently, the transmission spectrum develops a dense network of resonant channels that support efficient electron propagation across a broad energy range. In particular, the ordered system ($W=0$) exhibits a large number of sharp and well-defined resonances, especially at higher energies, reflecting the strong phase coherence of the underlying electronic states. The introduction of quasi-periodic modulation alters this coherent resonance structure through interference-induced spectral reconstruction, leading to a redistribution of transport-active states and a gradual reduction in the number and intensity of transmission peaks. The resulting evolution of the transmission spectrum highlights the delicate interplay between coupling-induced interference, quasi-periodic modulation, and lattice topology in determining the available conducting pathways.

The introduction of magnetic flux profoundly modifies the transmission characteristics in the dimerized regime ($t_1<t_2$), as illustrated in Fig.~\ref{fig:7}(c). By imparting an Aharonov--Bohm phase to the electronic wave function, the flux continuously reshapes the interference landscape of the ring and induces a substantial reconstruction of the transmission spectrum. As a consequence, numerous additional resonant channels emerge across a broad energy range, leading to a significantly richer distribution of transmission peaks and an overall enhancement of coherent transport. Despite this pronounced spectral reorganization, the suppression of transmission around the band center remains largely intact, indicating that the SSH dimerization gap is remarkably robust against flux-induced interference effects.

A particularly intriguing feature arises in the outer regions of the spectrum, where stronger quasi-periodic modulation produces a larger density of transmission resonances than the weakly modulated and ordered cases. This behavior reflects a nontrivial interplay between quasi-periodic disorder and magnetic-flux-induced quantum interference. Rather than acting solely as a scattering mechanism, the quasi-periodic potential cooperates with the flux-driven phase coherence to redistribute spectral weight and activate previously inaccessible resonant pathways. The resulting enhancement of transport-active states generates additional conducting channels in the high-energy sectors of the spectrum, highlighting the constructive role that disorder can play in shaping coherent transport under suitable interference conditions.

\subsubsection{Transmission driven transport response and current fluctuations}

For the homogeneous hopping configuration ($t_1=t_2$), the application of magnetic flux profoundly reshapes the transmission landscape, as shown in Fig.~\ref{fig:7}(d). The flux-induced Aharonov--Bohm phase reconstructs the underlying interference pattern, leading to a dense distribution of resonant transmission channels across the spectrum. Unlike the dimerized phase, finite transmission persists in the vicinity of the band center for all quasi-periodic modulation strengths, indicating the presence of transport-active states near the Fermi level. The simultaneous enhancement of low-energy transmission and the proliferation of resonant channels over a broad energy range substantially increase the number of states participating in transport. These observations demonstrate that hopping modulation, quasi-periodicity, and magnetic-flux-induced interference act collectively to redistribute transport-active states and thereby provide a powerful means of tailoring the transmission properties of the system.
\begin{figure*}[]
    \centering
    \includegraphics[width=15cm,height=4cm]{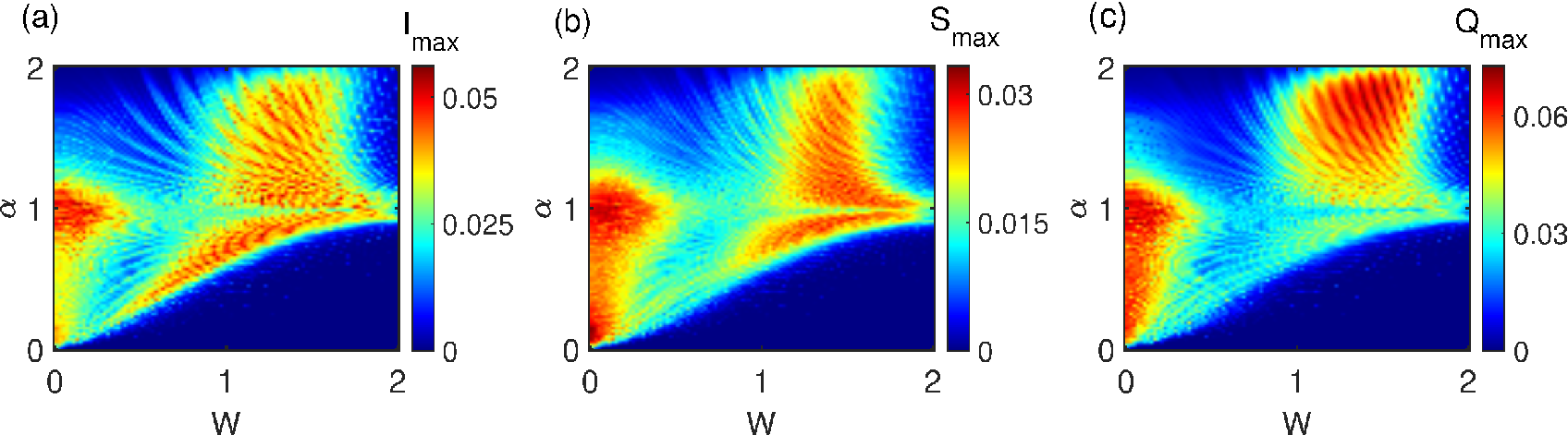}
    
\caption{
(Color online) Color map plots of the maximum (a) charge current, (b) current noise, and (c) heat current in the $\alpha$--$W$ parameter space for the asymmetric coupling geometry, where three lattice sites are connected to the source electrode and a single lattice site is connected to the drain electrode. The maxima are extracted within the bias-voltage window $1\leq V\leq2$. The magnetic flux is fixed at $\phi=0$. Other parameters are $\gamma=0.05,T_S=T+\Delta T$, $T_D=T-\Delta T$, $T=0.05$, $\Delta T=0.005$, $t_2=1$, $\alpha=t_1/t_2$, and $\mu_{S,D}=\mp V/2$.
}
\label{fig:9}
\end{figure*}
\begin{figure*}[]
    \centering
    \includegraphics[width=15cm,height=4cm]{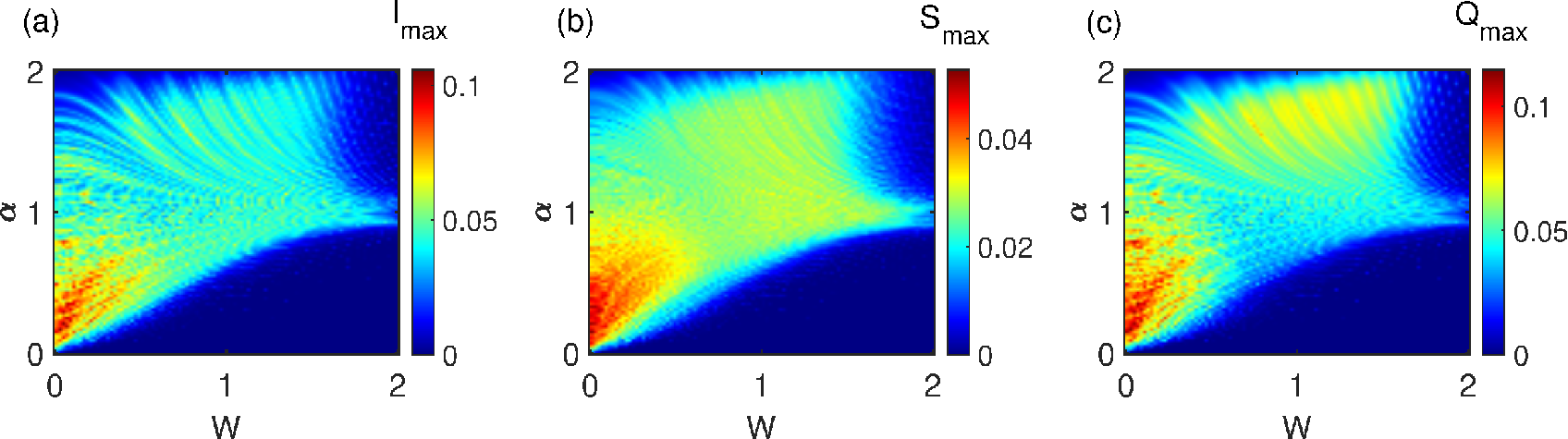}

\caption{(Color online) Color maps of the maximum (a) charge current, (b) current noise, and (c) heat current in the $\alpha$--$W$ parameter space for the asymmetric electrode-coupling configuration, where three lattice sites are connected to the source electrode and a single lattice site is connected to the drain electrode. The maximum values are extracted within the bias-voltage window $1\leq V\leq2$. Here, $\alpha=t_1/t_2$ denotes the degree of SSH dimerization and $W$ is the quasi-periodic modulation strength. Other parameters are identical to those in Fig.~\ref{fig:9}, except that $\phi=0.3$.
}
\label{fig:10}
\end{figure*}
The impact of this spectral restructuring on the transport characteristics is illustrated in Figs.~\ref{fig:8} and \ref{fig:15}, which show the bias dependence of the charge current, heat current, and current noise for the asymmetric configuration, where the source electrode is coupled to multiple lattice sites while the drain electrode is connected to a single site. Figure~\ref{fig:8} corresponds to the zero-flux configuration, whereas Fig.~\ref{fig:15} presents the corresponding results in the presence of magnetic flux. The left and right panels represent the dimerized ($t_1<t_2$) and homogeneous ($t_1=t_2$) hopping regimes, respectively.

In the absence of magnetic flux [Fig.~\ref{fig:8}], the transport response in the dimerized phase is strongest for the weakest quasi-periodic modulation and decreases monotonically with increasing disorder strength. This trend directly reflects the evolution of the transmission spectrum, where quasi-periodicity progressively suppresses the density and spectral weight of resonant conducting channels. As a result, fewer electronic states contribute within the bias window, leading to reduced charge transport, energy transport, and current fluctuations. Compared with the symmetric multi-site source--drain configuration discussed earlier, the overall magnitude of the transport response is substantially lower. This reduction originates from the asymmetric coupling geometry itself. Although electrons can enter the ring through multiple source channels, they must exit through a single drain site, creating an effective transport bottleneck that restricts the number of available conducting pathways. Consequently, the overall transmission probability is reduced, leading to a systematic suppression of all transport observables.

A markedly different transport behavior emerges in the homogeneous hopping regime ($t_1=t_2$), as shown in Figs.~\ref{fig:8}(d)--\ref{fig:8}(f). Owing to the absence of the SSH dimerization gap, finite charge current, heat current, and current fluctuations persist even in the low-bias region. This behavior can be directly traced to the transmission spectrum in Fig.~\ref{fig:7}(b), where a finite density of resonant conducting channels exists around the band center. The availability of these low-energy transport-active states enables sustained charge and energy transport over a broad bias range.

The influence of magnetic flux is illustrated in Fig.~\ref{fig:15}. By introducing an Aharonov--Bohm phase, the flux reshapes the underlying interference landscape and redistributes the transmission channels throughout the spectrum. In the dimerized regime ($t_1<t_2$), the transport response remains strongest for weak quasi-periodic modulation and decreases with increasing disorder strength. Although the flux activates additional resonant pathways, the dimerization-induced transport gap remains largely intact, thereby limiting the enhancement of transport. In contrast, the homogeneous hopping configuration exhibits a fundamentally different disorder dependence. Here, the strongest transport response is obtained for ($W=1$), reflecting a disorder-induced reconstruction of the transmission spectrum that generates a larger density of transport-active resonances within the bias window. Consequently, more electronic states participate in coherent transport, leading to enhanced charge and heat currents as well as larger current fluctuations.

A comparison between the zero- and finite-flux cases highlights the intricate interplay among quasi-periodicity, hopping modulation, and magnetic-flux-induced interference. In the dimerized phase, magnetic flux primarily enhances transport in the weakly disordered regime by activating additional coherent transmission pathways. In contrast, the homogeneous phase retains its disorder-assisted transport character even in the presence of flux, with the strongest modulation strength continuing to yield the largest transport response. These findings demonstrate that quasi-periodic disorder and magnetic flux do not act independently; rather, their cooperative and competing effects reshape the transmission landscape, providing a powerful mechanism for controlling coherent charge and energy transport.

\subsection{Global Transport Landscape in the $\alpha–W$ Parameter Space}
\begin{figure*}[t]
    \centering
    \includegraphics[width=15cm,height=4cm]{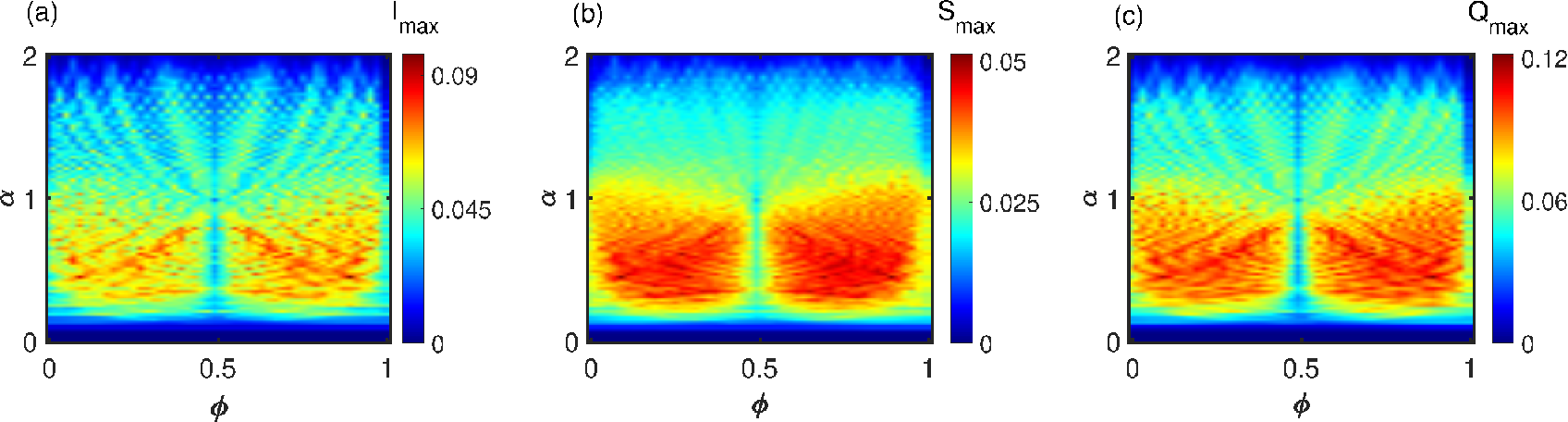}

\caption{
(Color online) Color map plots of the maximum (a) charge current, (b) current noise, and (c) heat current in the $\alpha$-$\phi$ parameter space for the symmetric coupling geometry. The maxima are extracted within the bias-voltage window $1\leq V\leq2$ at a fixed quasi-periodic modulation strength $W=0.3$. Here, $\alpha=t_1/t_2$ denotes the hopping-asymmetry parameter and $\phi$ is the magnetic flux. Other parameters are $\gamma=0.05,T_S=T+\Delta T$, $T_D=T-\Delta T$, $T=0.05$, $\Delta T=0.005$, $t_2=1$, and $\mu_{S,D}=\mp V/2$.
}
\label{fig:11}
\end{figure*}
\begin{figure*}[t]
    \centering
    \includegraphics[width=15cm,height=4cm]{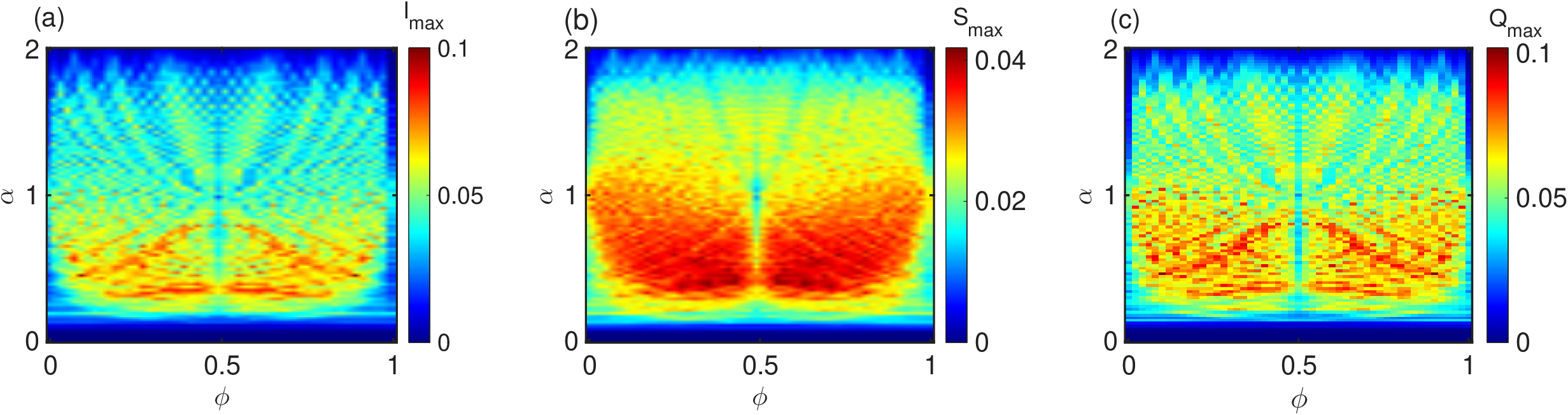}

\caption{(Color online) Color maps of the maximum (a) charge current, (b) current noise, and (c) heat current in the $\alpha$--$\phi$ parameter space for the asymmetric electrode-coupling configuration, where three lattice sites are connected to the source electrode and a single lattice site is connected to the drain electrode. The maximum values are extracted within the bias-voltage window $1\leq V\leq2$ at a fixed quasi-periodic modulation strength $W=0.3$. Here, $\alpha=t_1/t_2$ denotes the degree of SSH dimerization and $\phi$ is the magnetic flux. All other parameters are the same as those in Fig.~\ref{fig:11}.
}
\label{fig:12}
\end{figure*}
\begin{figure*}[t]
    \centering
    \includegraphics[width=15cm,height=4cm]{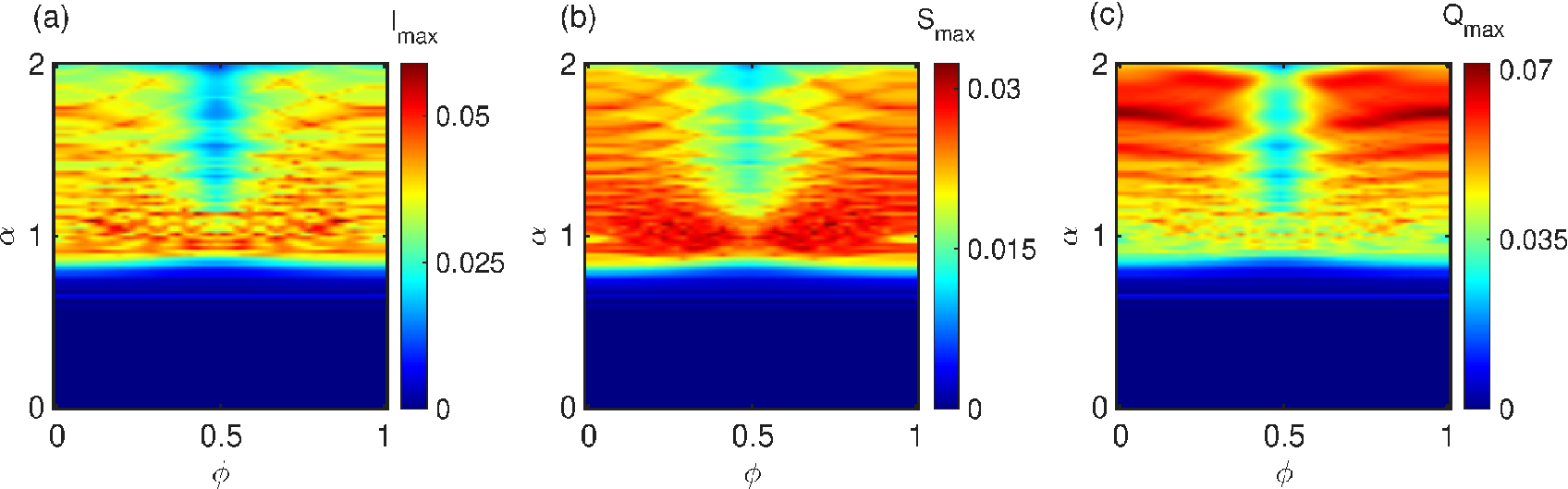}
   
\caption{
(Color online) Color map plots of the maximum (a) charge current, (b) current noise, and (c) heat current in the $\alpha$-$\phi$ parameter space for the asymmetric coupling geometry, where three lattice sites are coupled to the source electrode and a single lattice site is coupled to the drain electrode. The maxima are extracted within the bias-voltage window $1\leq V\leq2$ at a fixed quasi-periodic modulation strength $W=1.5$. Here, $\alpha=t_1/t_2$ denotes the hopping-asymmetry parameter and $\phi$ is the magnetic flux. Other parameters are $\gamma=0.05, T_S=T+\Delta T$, $T_D=T-\Delta T$, $T=0.05$, $\Delta T=0.005$, $t_2=1$, and $\mu_{S,D}=\mp V/2$.
}

    % \caption{Density plots of the maximum charge current, current noise, and heat current in the $\alpha-\phi$ parameter space for the asymmetric coupling geometry. The maxima are extracted over the bias-voltage window $1\leq V\leq2$ at a fixed disorder strength $W=1.5$. Other parameters are $T_S=T+\Delta T$, $T_D=T-\Delta T$, $T=0.05$, $\Delta T=0.005$, $t_2=1$, and $\mu_{S,D}=\mp V/2$.}
    \label{fig:16}
\end{figure*}

To obtain a global picture of the transport landscape, we map the maximum charge current, current noise, and heat current in the $(\alpha,W)$ parameter space, where $(\alpha=t_1/t_2)$ characterizes the degree of SSH dimerization and $(W)$ denotes the quasi-periodic modulation strength. The resulting phase diagrams for the symmetric and asymmetric coupling geometries are shown in Figs.~\ref{fig:4} and \ref{fig:5}, and Figs. \ref{fig:9}, and \ref{fig:10} respectively. The maxima are extracted within the bias window $1\leq V\leq2$, where the transport response is most pronounced.

For the symmetric multi-site source–drain configuration [Figs.~\ref{fig:4} and \ref{fig:5}], the transport-optimal regime is substantially extended beyond the homogeneous-hopping limit and deep into the topological dimerized phase $(\alpha<1)$. In contrast, for the conventional single-site coupling configuration, transport is maximized near the homogeneous-hopping limit $(\alpha \approx 1)$ and is strongly suppressed in the topological dimerized regime, as shown in Figs.~\ref{fig:13} and \ref{fig:14} of Appendix A. The extension of the high-transport region toward $(\alpha<1)$ demonstrates that multi-site electrode coupling fundamentally reshapes the underlying interference landscape and alters the hopping conditions required for optimal coherent transport. The application of magnetic flux [Fig.~9] further drives the optimal transport regime deeper into the topological phase while moderately enhancing the transport magnitude through the flux-induced emergence of additional resonant transmission pathways.

A qualitatively different phase diagram emerges for the asymmetric coupling geometry, where three sites are connected to the source electrode and a single site to the drain [Figs.~\ref{fig:9} and ~\ref{fig:10}]. In the absence of magnetic flux (Fig.~\ref{fig:9}), the high-transport region extends over a broader range of $(\alpha)$, and a pronounced re-entrant transport phase appears at intermediate and large quasiperiodic modulation strengths. This disorder-assisted enhancement is absent in both conventional single-site geometries and the symmetric multi-site configuration, indicating that it originates from the redistribution of quantum-interference pathways induced by coupling asymmetry. In this regime, quasiperiodicity no longer acts solely as a transport-suppressing mechanism but instead promotes transport through the formation of additional resonant conducting channels.

The application of magnetic flux substantially modifies this transport landscape [Fig.~\ref{fig:10}]. Although the overall transport response is enhanced, the disorder-assisted high-transport region observed at larger $(W)$ is strongly suppressed, and the optimal transport window shifts back toward the dimerized regime $(\alpha<1)$. This behavior reveals a competition between two distinct transport-enhancement mechanisms: disorder-assisted resonant conduction driven by coupling asymmetry and interference-controlled transport governed by the Aharonov--Bohm phase.

Taken together, Figs.~\ref{fig:4}, \ref{fig:5}, \ref{fig:9}, and \ref{fig:10} establish electrode-coupling engineering as a powerful control parameter for nonequilibrium transport in quasiperiodic systems. By modifying the reservoir-coupling architecture, one can not only relocate the transport-optimal regime within the $(\alpha,W)$ phase space but also induce or suppress entirely new transport phases arising from the interplay of dimerization, quasiperiodicity, and magnetic-flux-induced quantum interference.
\subsection{Global Transport Landscape in the $\alpha–\phi$ Parameter Space}

To further elucidate the role of magnetic-flux-induced interference, Figs.~\ref{fig:11}, \ref{fig:12} , and \ref{fig:16} respectively, present the maximum charge current, heat current, and current noise in the $(\alpha,\phi)$ parameter space for different coupling geometries and quasiperiodic modulation strengths. These maps provide a global view of how hopping dimerization and magnetic flux jointly determine the transport response.

For weak quasiperiodic modulation ($W=0.3$), the symmetric multi-site coupling geometry (Fig.~\ref{fig:11}) exhibits a well-defined high-transport region predominantly confined to the dimerized regime $(\alpha<1)$. This behavior reflects the strong influence of contact-engineered interference pathways, which shift the transport-optimal regime away from the conventional homogeneous-hopping limit. Although magnetic flux modifies the interference conditions through the Aharonov--Bohm phase, the overall structure of the transport landscape remains largely intact, indicating that the favorable transport regime is primarily dictated by the coupling geometry.

A qualitatively different behavior emerges for the asymmetric coupling configuration (Fig.~\ref{fig:12}). Here, the high-transport region extends over a broader range of $(\alpha)$, and additional transport-favorable domains appear due to the redistribution of transmission channels induced by coupling asymmetry. The resulting transport enhancement highlights the ability of asymmetric contacts to activate interference-assisted conducting pathways that are absent in the symmetric geometry.

The influence of quasiperiodicity becomes particularly pronounced at stronger modulation strengths ($W=1.5$), as shown in Fig.~\ref{fig:16}. In this regime, the transport-optimal window progressively shifts away from the topological dimerized phase and toward the vicinity of the trivial-hopping regime ($\alpha \geq 1$). This disorder-driven migration reflects the gradual reduction of the transport advantage associated with strong dimerization and signals the emergence of more favorable transport conditions in nearly uniform lattices. While magnetic flux continues to modulate the magnitude of the transport response, the location of the optimal transport window is governed primarily by the competition between SSH dimerization and quasiperiodicity-induced spectral reconstruction.

Taken together, Figs.~\ref{fig:11}, \ref{fig:12}, and \ref{fig:16} demonstrate that the transport landscape is shaped by the intricate interplay among hopping modulation, quasiperiodicity, magnetic-flux-induced interference, and contact-engineered coupling geometry. Most significantly, increasing quasiperiodic modulation induces a systematic migration of the transport-optimal regime from the topological dimerized phase toward $(\alpha \geq 1)$, highlighting the possibility of engineering coherent transport through the cooperative action of disorder, quantum interference, and reservoir-coupling design.

\begin{figure*}[t]
    \centering
    \includegraphics[width=15cm,height=4cm]{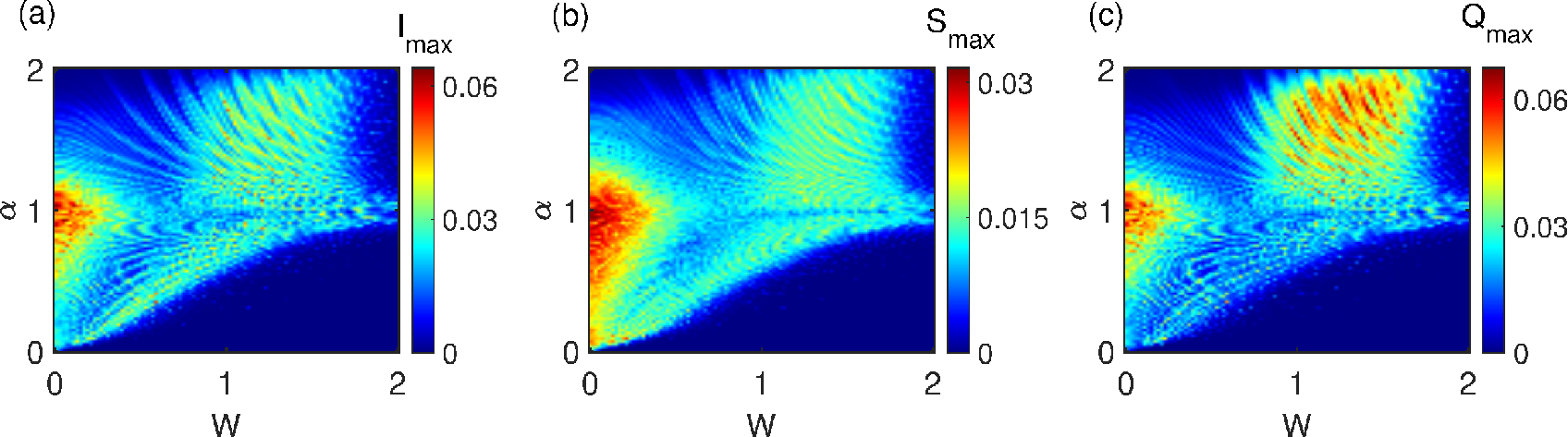}

\caption{
(Color online) colormap plots of the maximum (a) charge current, (b) current noise, and (c) heat current in the $\alpha$--$W$ parameter space for the ring geometry with single-site source and drain coupling. The maxima are extracted within the bias-voltage window $1\leq V\leq2$. The magnetic flux is fixed at $\phi=0$. Here, $\alpha=t_1/t_2$ characterizes the degree of SSH dimerization and $W$ denotes the quasiperiodic modulation strength. Other parameters are $\gamma=0.05, T_S=T+\Delta T$, $T_D=T-\Delta T$, $T=0.05$, $\Delta T=0.005$, $t_2=1$, and $\mu_{S,D}=\mp V/2$.
}
\label{fig:13}
\end{figure*}
\begin{figure*}[t]
    \centering
    \includegraphics[width=15cm,height=4cm]{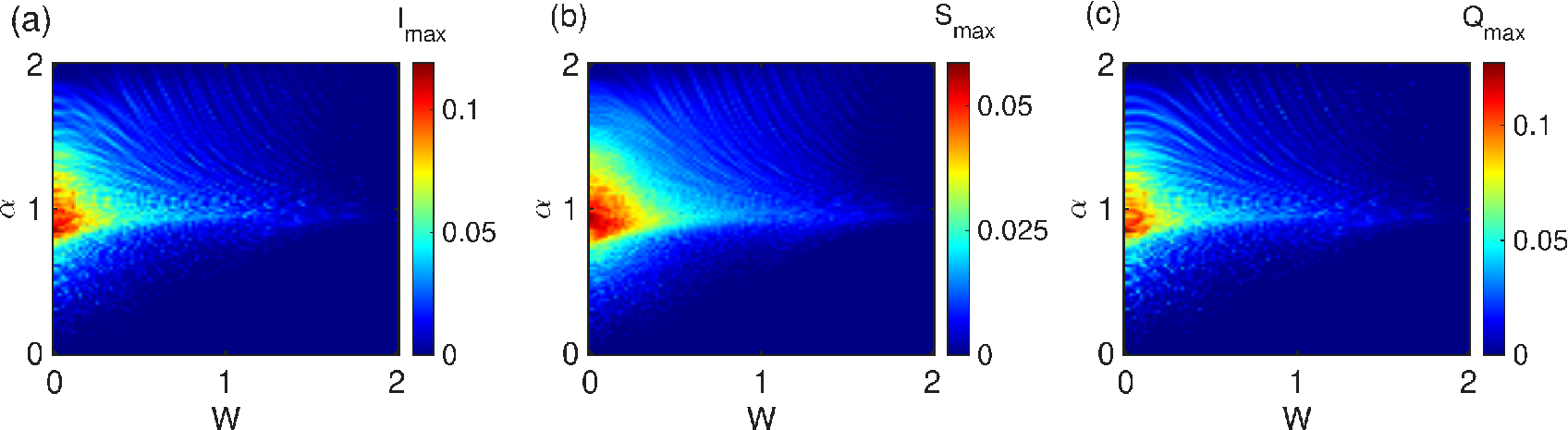}
   \caption{
(Color online) colormap plots of the maximum (a) charge current, (b) current noise, and (c) heat current in the $\alpha$--$W$ parameter space for the open-chain geometry with single-site source and drain coupling. The maxima are extracted within the bias-voltage window $1\leq V\leq2$. Here, $\alpha=t_1/t_2$ denotes the hopping-asymmetry parameter and $W$ is the quasiperiodic modulation strength. Other parameters are $\gamma=0.05,T_S=T+\Delta T$, $T_D=T-\Delta T$, $T=0.05$, $\Delta T=0.005$, $t_2=1$, and $\mu_{S,D}=\mp V/2$.
}
\label{fig:14}
\end{figure*}
\subsection{Experimental realization and future perspectives}
The contact-engineering mechanism proposed here is experimentally accessible in several quantum platforms. Gate-defined semiconductor quantum-dot arrays provide a particularly direct realization, where individual dots act as lattice sites and electrostatic gates allow source and drain reservoirs to be coupled to one or multiple neighboring dots. Such tunable coupling architectures naturally realize the symmetric and asymmetric multi-site contact configurations considered in this work, while ring geometries enable the incorporation of magnetic-flux-induced Aharonov--Bohm interference. Similar contact-dependent transport phenomena may also be explored in molecular junctions, where electrodes can couple to multiple molecular orbitals through different anchoring groups, as well as in artificial atomic lattices and photonic SSH systems with engineered boundary couplings. The predicted transport signatures, including disorder-assisted transport and flux-controlled modulation of conducting phases, should therefore be accessible with current experimental capabilities. These considerations elevate contact architecture from a passive boundary condition to an active functional element of quantum-device design, opening new avenues for manipulating charge and energy transport through the deliberate engineering of interference pathways at the nanoscale.

The ability to engineer transport characteristics through the combined action of quasiperiodicity, quantum interference, magnetic flux, and tailored contact architectures opens new opportunities for controlling charge and energy flow in mesoscopic systems. More broadly, our findings establish contact engineering as a versatile strategy for optimizing coherent transport and thermoelectric performance in low-dimensional quantum devices, providing a promising framework for the development of next-generation energy-harvesting and quantum-functional materials.
\section{Conclusion}\label{secIV}
In this work, we have demonstrated that the geometry of system-reservoir coupling can fundamentally reshape coherent transport in quasiperiodic quantum systems. By investigating charge transport, heat transport, and current fluctuations in a magnetic-flux-threaded Aubry--Andr\'e--Harper-modulated SSH ring within the nonequilibrium Green's function framework, we have uncovered a rich transport landscape governed by the intricate interplay of hopping dimerization, quasiperiodic modulation, magnetic-flux-induced quantum interference, and contact architecture.

Our results reveal that contact-engineered coupling is not merely a boundary condition but an active control parameter capable of restructuring the distribution of transport-active states. While SSH dimerization opens a transport gap and suppresses conduction near the band center, magnetic flux reconstructs the interference landscape through the Aharonov--Bohm effect, generating additional resonant conducting channels. More remarkably, multi-site electrode coupling qualitatively reshapes the conditions for optimal transport, extending the transport-favorable regime far beyond that observed in conventional single-site coupling geometries. The location of the transport-optimal window is found to be highly sensitive to the coupling geometry, occurring either deep within the topological dimerized phase or in the vicinity of the trivial-hopping regime. This tunability highlights the central role of contact engineering in controlling coherent transport. The resulting transport characteristics arise from the intricate interplay among quasiperiodicity-induced spectral reconstruction, quantum interference, and localization effects.

A central finding of this work is the emergence of disorder-assisted transport under asymmetric coupling conditions. In contrast to the conventional expectation that increasing disorder suppresses coherent transport, we find that moderate quasiperiodic modulation can activate additional resonant pathways and enhance both charge and energy transfer. This effect originates from a coupling-induced redistribution of interference pathways that transforms quasiperiodicity from a purely transport-suppressing mechanism into a resource for coherent conduction. Furthermore, increasing the quasiperiodic modulation strength progressively relocates the transport-optimal window from the topological dimerized regime toward the vicinity of the trivial-hopping limit, highlighting the intricate competition among localization, dimerization, and interference effects in determining the optimal transport conditions. Collectively, these findings establish contact engineering as a powerful route for controlling coherent transport in quasiperiodic quantum systems and provide a versatile framework for designing transport functionalities through the combined manipulation of coupling geometry, disorder, and quantum interference.

\appendix

\section{Justification for multi-site system-electrode coupling}
Before discussing the transport properties of the multi-site coupled geometries considered in the main text, it is instructive to examine the corresponding behavior of the conventional configurations in which the source and drain electrodes are connected to the system through a single lattice site. To this end, in Figs.~\ref{fig:13} and \ref{fig:14} we present the maximum charge current, heat current, and current noise in the $\alpha$--$W$ parameter space for the ring and chain geometries, respectively.

A common feature of both geometries is that the transport response is maximized in the vicinity of the homogeneous hopping condition, $\alpha=t_1/t_2\approx1$. As the hopping amplitudes become increasingly unequal, the transport efficiency decreases substantially due to the enhanced dimerization of the lattice. Consequently, the transport-favorable region remains localized around $\alpha\approx1$ over a broad range of disorder strengths. Although the quasiperiodic potential modifies the magnitude of the transport coefficients, it does not qualitatively alter the location of the optimal transport regime.

These observations establish the conventional behavior expected for single-site coupled quasiperiodic SSH systems and provide a natural benchmark against which the multi-site coupling geometries studied in the main text can be compared. Remarkably, once multiple lattice sites are connected to the source and/or drain electrodes, the transport landscape undergoes a qualitative transformation. The region of maximal transport is no longer restricted to the homogeneous hopping regime and instead shifts toward the dimerized phase ($\alpha<1$). Furthermore, asymmetric multi-site coupling generates additional transport-favorable regions and disorder-assisted transport enhancement that are entirely absent in the conventional single-site coupled chain and ring geometries.

Therefore, the comparison demonstrates that the unconventional transport features reported in the main text are not an intrinsic consequence of quasiperiodicity alone. Rather, they originate from the profound modification of quantum interference pathways induced by multi-site electrode coupling. The coupling engineering effectively redistributes the available transmission channels and changes the conditions under which transport is optimized. This provides the primary motivation for considering multi-site coupling configurations and highlights coupling engineering as an efficient mechanism for manipulating charge and energy transport in quasiperiodic mesoscopic systems.

\bibliography{references}

@article{l9gd-k9yw,
  title = {Coherent control of thermoelectric performance via engineered transmission functions in multidot Aharonov-Bohm heat engines},
  author = {Sridhar and Bedkihal, Salil and Bandyopadhyay, Malay},
  journal = {Phys. Rev. B},
  volume = {113},
  issue = {8},
  pages = {085428},
  numpages = {20},
  year = {2026},
  month = {Feb},
  publisher = {American Physical Society},
  doi = {10.1103/l9gd-k9yw},
  url = {https://link.aps.org/doi/10.1103/l9gd-k9yw}
}

@article{dhar2006nonequilibrium,
  title={Nonequilibrium Green’s function formalism and the problem of bound states},
  author={Dhar, Abhishek and Sen, Diptiman},
  journal={Physical Review B—Condensed Matter and Materials Physics},
  volume={73},
  number={8},
  pages={085119},
  year={2006},
  publisher={APS}
}

@book{datta1997electronic,
  title={Electronic transport in mesoscopic systems},
  author={Datta, Supriyo},
  year={1997},
  publisher={Cambridge university press}
}

@article{yamamoto2015thermodynamics,
  title={Thermodynamics of the mesoscopic thermoelectric heat engine beyond the linear-response regime},
  author={Yamamoto, Kaoru and Hatano, Naomichi},
  journal={Physical Review E},
  volume={92},
  number={4},
  pages={042165},
  year={2015},
  publisher={APS}
}

@article{sivan1986multichannel,
  title={Multichannel Landauer formula for thermoelectric transport with application to thermopower near the mobility edge},
  author={Sivan, U and Imry, Y},
  journal={Physical review b},
  volume={33},
  number={1},
  pages={551},
  year={1986},
  publisher={APS}
}

@article{butcher1990thermal,
  title={Thermal and electrical transport formalism for electronic microstructures with many terminals},
  author={Butcher, PN},
  journal={Journal of Physics: Condensed Matter},
  volume={2},
  number={22},
  pages={4869},
  year={1990},
  publisher={IOP Publishing}
}

@article{benenti2017fundamental,
  title={Fundamental aspects of steady-state conversion of heat to work at the nanoscale},
  author={Benenti, Giuliano and Casati, Giulio and Saito, Keiji and Whitney, Robert S},
  journal={Physics Reports},
  volume={694},
  pages={1--124},
  year={2017},
  publisher={Elsevier}
}

@article{yang2018gate,
  title={Gate voltage controlled thermoelectric figure of merit in three-dimensional topological insulator nanowires},
  author={Yang, Ning-Xuan and Zhou, Yan-Feng and Lv, Peng and Sun, Qing-Feng},
  journal={Physical Review B},
  volume={97},
  number={23},
  pages={235435},
  year={2018},
  publisher={APS}
}

@article{esposito2015quantum,
  title={Quantum thermodynamics: A nonequilibrium Green’s function approach},
  author={Esposito, Massimiliano and Ochoa, Maicol A and Galperin, Michael},
  journal={Physical review letters},
  volume={114},
  number={8},
  pages={080602},
  year={2015},
  publisher={APS}
}

@article{esposito2015nature,
  title={Nature of heat in strongly coupled open quantum systems},
  author={Esposito, Massimiliano and Ochoa, Maicol A and Galperin, Michael},
  journal={Physical Review B},
  volume={92},
  number={23},
  pages={235440},
  year={2015},
  publisher={APS}
}

@article{seshadri2021entropy,
  title={Entropy and information flow in quantum systems strongly coupled to baths},
  author={Seshadri, Nikhil and Galperin, Michael},
  journal={Physical Review B},
  volume={103},
  number={8},
  pages={085415},
  year={2021},
  publisher={APS}
}

@article{bergmann2021green,
  title={A Green’s function perspective on the nonequilibrium thermodynamics of open quantum systems strongly coupled to baths: Nonequilibrium quantum thermodynamics},
  author={Bergmann, Nicolas and Galperin, Michael},
  journal={The European Physical Journal Special Topics},
  volume={230},
  number={4},
  pages={859--866},
  year={2021},
  publisher={Springer}
}

@article{topp2015steady,
  title={Steady-state thermodynamics of non-interacting transport beyond weak coupling},
  author={Topp, Gabriel E and Brandes, Tobias and Schaller, Gernot},
  journal={Europhysics Letters},
  volume={110},
  number={6},
  pages={67003},
  year={2015},
  publisher={IOP Publishing}
}

@article{mazza2014thermoelectric,
  title={Thermoelectric efficiency of three-terminal quantum thermal machines},
  author={Mazza, Francesco and Bosisio, Riccardo and Benenti, Giuliano and Giovannetti, Vittorio and Fazio, Rosario and Taddei, Fabio},
  journal={New Journal of Physics},
  volume={16},
  number={8},
  pages={085001},
  year={2014},
  publisher={IOP Publishing}
}

@article{PhysRevB.98.155438,
  title = {Assessing the validity of the thermodynamic uncertainty relation in quantum systems},
  author = {Agarwalla, Bijay Kumar and Segal, Dvira},
  journal = {Phys. Rev. B},
  volume = {98},
  issue = {15},
  pages = {155438},
  numpages = {9},
  year = {2018},
  month = {Oct},
  publisher = {American Physical Society},
  doi = {10.1103/PhysRevB.98.155438},
  url = {https://link.aps.org/doi/10.1103/PhysRevB.98.155438}
}

@article{PhysRevE.99.062141,
  title = {Thermodynamic uncertainty relation in quantum thermoelectric junctions},
  author = {Liu, Junjie and Segal, Dvira},
  journal = {Phys. Rev. E},
  volume = {99},
  issue = {6},
  pages = {062141},
  numpages = {12},
  year = {2019},
  month = {Jun},
  publisher = {American Physical Society},
  doi = {10.1103/PhysRevE.99.062141},
  url = {https://link.aps.org/doi/10.1103/PhysRevE.99.062141}
}

@article{PhysRevE.100.042101,
  title = {Thermodynamic uncertainty relation in thermal transport},
  author = {Saryal, Sushant and Friedman, Hava Meira and Segal, Dvira and Agarwalla, Bijay Kumar},
  journal = {Phys. Rev. E},
  volume = {100},
  issue = {4},
  pages = {042101},
  numpages = {10},
  year = {2019},
  month = {Oct},
  publisher = {American Physical Society},
  doi = {10.1103/PhysRevE.100.042101},
  url = {https://link.aps.org/doi/10.1103/PhysRevE.100.042101}
}

@article{AubryAndre1980,
  author  = {S. Aubry and G. Andr\'e},
  title   = {Analyticity breaking and Anderson localization in incommensurate lattices},
  journal = {Ann. Isr. Phys. Soc.},
  volume  = {3},
  pages   = {18},
  year    = {1980}
}

@article{Harper1955,
  author  = {P. G. Harper},
  title   = {The General Motion of Conduction Electrons in a Uniform Magnetic Field, with Application to the Diamagnetism of Metals},
  journal = {Proc. Phys. Soc. A},
  volume  = {68},
  pages   = {874},
  year    = {1955},
  doi     = {10.1088/0370-1298/68/10/304}
}

@article{Su1979,
  author  = {W. P. Su and J. R. Schrieffer and A. J. Heeger},
  title   = {Solitons in Polyacetylene},
  journal = {Physical Review Letters},
  volume  = {42},
  pages   = {1698--1701},
  year    = {1979},
  doi     = {10.1103/PhysRevLett.42.1698}
}

@article{Su1980,
  author  = {W. P. Su and J. R. Schrieffer and A. J. Heeger},
  title   = {Soliton Excitations in Polyacetylene},
  journal = {Physical Review B},
  volume  = {22},
  pages   = {2099--2111},
  year    = {1980},
  doi     = {10.1103/PhysRevB.22.2099}
}

@article{Asboth2016,
  author    = {J. K. Asb{\'o}th and L. Oroszl{\'a}ny and A. P{\'a}lyi},
  title     = {A Short Course on Topological Insulators: Band Structure and Edge States in One and Two Dimensions},
  journal   = {Lecture Notes in Physics},
  volume    = {919},
  pages     = {1--166},
  year      = {2016},
  publisher = {Springer},
  doi       = {10.1007/978-3-319-25607-8}
}

@article{Ryu2002,
  author  = {S. Ryu and Y. Hatsugai},
  title   = {Topological Origin of Zero-Energy Edge States in Particle-Hole Symmetric Systems},
  journal = {Physical Review Letters},
  volume  = {89},
  pages   = {077002},
  year    = {2002},
  doi     = {10.1103/PhysRevLett.89.077002}
}

@article{Deng2019,
  author  = {Xiang Deng and S. Ray and S. Sinha and G. V. Shlyapnikov and L. Santos},
  title   = {Reentrant Localization Transition in a Quasiperiodic Chain},
  journal = {Physical Review Letters},
  volume  = {123},
  pages   = {025301},
  year    = {2019},
  doi     = {10.1103/PhysRevLett.123.025301}
}

@article{Mazza2014,
  author  = {L. Mazza and J. V. Balachandran and P. Simon and R. Egger},
  title   = {Thermoelectric performance of topological boundary modes},
  journal = {Physical Review B},
  volume  = {91},
  pages   = {245435},
  year    = {2015},
  doi     = {10.1103/PhysRevB.91.245435}
}

@book{Datta1995,
  author    = {S. Datta},
  title     = {Electronic Transport in Mesoscopic Systems},
  publisher = {Cambridge University Press},
  address   = {Cambridge},
  year      = {1995}
}

@book{Datta2005,
  author    = {S. Datta},
  title     = {Quantum Transport: Atom to Transistor},
  publisher = {Cambridge University Press},
  address   = {Cambridge},
  year      = {2005}
}

@book{Imry2002,
  author    = {Y. Imry},
  title     = {Introduction to Mesoscopic Physics},
  edition   = {2},
  publisher = {Oxford University Press},
  address   = {Oxford},
  year      = {2002}
}

@article{Landauer1970,
  author  = {R. Landauer},
  title   = {Electrical Resistance of Disordered One-Dimensional Lattices},
  journal = {Philosophical Magazine},
  volume  = {21},
  number  = {172},
  pages   = {863--867},
  year    = {1970}
}

@article{Buttiker1986,
  author  = {M. B{\"u}ttiker},
  title   = {Four-Terminal Phase-Coherent Conductance},
  journal = {Physical Review Letters},
  volume  = {57},
  number  = {14},
  pages   = {1761--1764},
  year    = {1986},
  doi     = {10.1103/PhysRevLett.57.1761}
}

@article{Roati2008,
  author  = {G. Roati and C. D'Errico and L. Fallani and M. Fattori and C. Fort and M. Zaccanti and G. Modugno and M. Modugno and M. Inguscio},
  title   = {Anderson Localization of a Non-Interacting Bose–Einstein Condensate},
  journal = {Nature},
  volume  = {453},
  pages   = {895--898},
  year    = {2008},
  doi     = {10.1038/nature07071}
}

@article{Lahini2009,
  author  = {Y. Lahini and A. Avidan and F. Pozzi and M. Sorel and R. Morandotti and D. N. Christodoulides and Y. Silberberg},
  title   = {Observation of a Localization Transition in Quasiperiodic Photonic Lattices},
  journal = {Physical Review Letters},
  volume  = {103},
  pages   = {013901},
  year    = {2009},
  doi     = {10.1103/PhysRevLett.103.013901}
}

@article{Kraus2012,
  author  = {Y. E. Kraus and Y. Lahini and Z. Ringel and M. Verbin and O. Zilberberg},
  title   = {Topological States and Adiabatic Pumping in Quasicrystals},
  journal = {Physical Review Letters},
  volume  = {109},
  pages   = {106402},
  year    = {2012},
  doi     = {10.1103/PhysRevLett.109.106402}
}

@article{Meier2016,
  author  = {E. J. Meier and F. A. An and B. Gadway},
  title   = {Observation of the topological soliton state in the Su–Schrieffer–Heeger model},
  journal = {Nature Communications},
  volume  = {7},
  pages   = {13986},
  year    = {2016},
  doi     = {10.1038/ncomms13986}
}

@article{Aharonov1959,
  author  = {Y. Aharonov and D. Bohm},
  title   = {Significance of Electromagnetic Potentials in the Quantum Theory},
  journal = {Physical Review},
  volume  = {115},
  number  = {3},
  pages   = {485--491},
  year    = {1959},
  doi     = {10.1103/PhysRev.115.485}
}

@article{Buttiker1983,
  author  = {M. Büttiker and Y. Imry and M. Ya. Azbel},
  title   = {Quantum oscillations in one-dimensional normal-metal rings},
  journal = {Physics Letters A},
  volume  = {96},
  number  = {7},
  pages   = {365--367},
  year    = {1983},
  doi     = {10.1016/0375-9601(83)90011-7}
}

@article{Gefen1984,
  author  = {Y. Gefen and Y. Imry and M. Ya. Azbel},
  title   = {Quantum Oscillations and the Aharonov-Bohm Effect for Parallel Resistors},
  journal = {Physical Review Letters},
  volume  = {52},
  number  = {2},
  pages   = {129--132},
  year    = {1984},
  doi     = {10.1103/PhysRevLett.52.129}
}

@article{Bedkihal2025,
  author  = {Salil Bedkihal and Jayasmita Behera and Malay Bandyopadhyay},
  title   = {Fundamental aspects of Aharonov--Bohm quantum machines: thermoelectric heat engines and diodes},
  journal = {Journal of Physics: Condensed Matter},
  volume  = {37},
  number  = {16},
  pages   = {163001--163036},
  year    = {2025},
  doi     = {10.1088/1361-648X/adb921}
}

@article{Aydin2025,
  author  = {A. Aydin and S. Bedkihal and M. Bandyopadhyay},
  title   = {Geometry-induced asymmetric level coupling},
  journal = {Physical Review E},
  volume  = {112},
  pages   = {014121},
  year    = {2025},
  doi     = {10.1103/PhysRevE.112.014121}
}

@article{Sitek2013,
  author  = {A. Sitek and A. Manolescu},
  title   = {Dicke states in multiple quantum dots},
  journal = {Physical Review A},
  volume  = {88},
  pages   = {043807},
  year    = {2013},
  doi     = {10.1103/PhysRevA.88.043807}
}

@article{Dicke1953,
  author  = {R. H. Dicke},
  title   = {The Effect of Collisions upon the Doppler Width of Spectral Lines},
  journal = {Physical Review},
  volume  = {89},
  number  = {2},
  pages   = {472--473},
  year    = {1953},
  doi     = {10.1103/PhysRev.89.472}
}

@article{Wang2013,
  author  = {Q. Wang and H. Xie and Y.-H. Nie and W. Ren},
  title   = {Enhancement of thermoelectric efficiency in triple quantum dots by the Dicke effect},
  journal = {Physical Review B},
  volume  = {87},
  pages   = {075102},
  year    = {2013},
  doi     = {10.1103/PhysRevB.87.075102}
}

@article{Liu2015,
  author  = {F. Liu and S. Ghosh and Y. D. Chong},
  title   = {Localization and adiabatic pumping in a generalized Aubry-Andr\'e-Harper model},
  journal = {Physical Review B},
  volume  = {91},
  pages   = {014108},
  year    = {2015},
  doi     = {10.1103/PhysRevB.91.014108}
}

@article{Ganeshan2015,
  author  = {S. Ganeshan and J. H. Pixley and S. Das Sarma},
  title   = {Nearest neighbor tight binding models with an exact mobility edge in one dimension},
  journal = {Physical Review Letters},
  volume  = {114},
  pages   = {146601},
  year    = {2015},
  doi     = {10.1103/PhysRevLett.114.146601}
}

@article{Wang2020,
  author  = {Y. Wang and X. Li and J. Gong},
  title   = {Thermoelectric transport in quasiperiodic topological systems},
  journal = {Physical Review B},
  volume  = {101},
  pages   = {085121},
  year    = {2020},
  doi     = {10.1103/PhysRevB.101.085121}
}

@article{Bhattacharya2025,
  author  = {Ranjini Bhattacharya and Souvik Roy},
  title   = {Fibonacci-engineered spin and charge thermoelectrics in a long range Su--Schrieffer--Heeger chain: A pathway to giant figure of merit},
  journal = {Journal of Applied Physics},
  volume  = {138},
  pages   = {184301},
  year    = {2025},
  doi     = {10.1063/5.0297435}
}

@article{xv5t-qvcm,
  title = {Weak localization and universal conductance fluctuations in large-area twisted bilayer graphene},
  author = {Talkington, Spenser and Mallick, Debarghya and Chen, An-Hsi and Mead, Benjamin F. and Yang, Seong-Jun and Kim, Cheol-Joo and Adam, Shaffique and Wu, Liang and Brahlek, Matthew and Mele, Eugene J.},
  journal = {Phys. Rev. B},
  volume = {113},
  issue = {16},
  pages = {165430},
  numpages = {10},
  year = {2026},
  month = {Apr},
  publisher = {American Physical Society},
  doi = {10.1103/xv5t-qvcm},
  url = {https://link.aps.org/doi/10.1103/xv5t-qvcm}
}

@article{PhysRevB.97.174206,
  title = {Anomalous transport in the Aubry-Andr\'e-Harper model in isolated and open systems},
  author = {Purkayastha, Archak and Sanyal, Sambuddha and Dhar, Abhishek and Kulkarni, Manas},
  journal = {Phys. Rev. B},
  volume = {97},
  issue = {17},
  pages = {174206},
  numpages = {11},
  year = {2018},
  month = {May},
  publisher = {American Physical Society},
  doi = {10.1103/PhysRevB.97.174206},
  url = {https://link.aps.org/doi/10.1103/PhysRevB.97.174206}
}

@article{article,
author = {Domaretskiy, Daniil and Wu, Zefei and Nguyen, Van Huy and Hayward, Ned and Babich, Ian and Li, Xiao and Nguyen, Ekaterina and Barrier, Julien and Indykiewicz, Kornelia and Wang, Wendong and Gorbachev, Roman and Xin, Na and Watanabe, Kenji and Taniguchi, Takashi and Hague, Lee and Falko, Vladimir and Grigorieva, Irina and Ponomarenko, Leonid and Berdyugin, Alexey and Geim, Andre},
year = {2025},
month = {08},
pages = {646-651},
title = {Proximity screening greatly enhances electronic quality of graphene},
volume = {644},
journal = {Nature},
doi = {10.1038/s41586-025-09386-0}
}

@article{Deng,
author = {Deng, Peng and Zhang, Peng and Qiu, Gang and Yang, Ting‐Hsun and Niu, Chang and Li, Yaochen and Cui, Wenqiang and Feng, Yang and Ye, Peide and He, Ke and Chang, Kai and Wang, Kang},
year = {2025},
month = {12},
pages = {},
title = {Universal Conductance Fluctuations in Quantum Anomalous Hall Insulators},
journal = {Advanced Materials},
doi = {10.1002/adma.202518012}
}

@article{Yildiz2025,
  author  = {Taylan Yildiz and B. Tanatar},
  title   = {Localization and persistent currents in a quasiperiodic disordered helical lattice},
  journal = {Scientific Reports},
  volume  = {15},
  number  = {1},
  pages   = {37307},
  year    = {2025},
  doi     = {10.1038/s41598-025-21294-x},
  issn    = {2045-2322},
  url     = {https://doi.org/10.1038/s41598-025-21294-x}
}

@article{Maiti2011,
  author  = {Santanu K. Maiti},
  title   = {Determination of Rashba and Dresselhaus spin-orbit fields},
  journal = {Journal of Applied Physics},
  volume  = {110},
  number  = {6},
  pages   = {064306},
  year    = {2011},
  doi     = {10.1063/1.3632060}
}

@article{PhysRevLett.109.116404,
  title = {Topological Equivalence between the Fibonacci Quasicrystal and the Harper Model},
  author = {Kraus, Yaacov E. and Zilberberg, Oded},
  journal = {Phys. Rev. Lett.},
  volume = {109},
  issue = {11},
  pages = {116404},
  numpages = {5},
  year = {2012},
  month = {Sep},
  publisher = {American Physical Society},
  doi = {10.1103/PhysRevLett.109.116404},
  url = {https://link.aps.org/doi/10.1103/PhysRevLett.109.116404}
}

@article{PhysRevLett.114.146601,
  title = {Nearest Neighbor Tight Binding Models with an Exact Mobility Edge in One Dimension},
  author = {Ganeshan, Sriram and Pixley, J. H. and Das Sarma, S.},
  journal = {Phys. Rev. Lett.},
  volume = {114},
  issue = {14},
  pages = {146601},
  numpages = {5},
  year = {2015},
  month = {Apr},
  publisher = {American Physical Society},
  doi = {10.1103/PhysRevLett.114.146601},
  url = {https://link.aps.org/doi/10.1103/PhysRevLett.114.146601}
}

@article{PhysRevB.87.045418,
  title = {Flux-dependent occupations and occupation difference in geometrically symmetric and energy degenerate double-dot Aharonov-Bohm interferometers},
  author = {Bedkihal, Salil and Bandyopadhyay, Malay and Segal, Dvira},
  journal = {Phys. Rev. B},
  volume = {87},
  issue = {4},
  pages = {045418},
  numpages = {12},
  year = {2013},
  month = {Jan},
  publisher = {American Physical Society},
  doi = {10.1103/PhysRevB.87.045418},
  url = {https://link.aps.org/doi/10.1103/PhysRevB.87.045418}
}

@article{PhysRevB.83.115318,
  title = {Intrinsic coherence dynamics and phase localization in nanoscale Aharonov-Bohm interferometers},
  author = {Tu, Matisse Wei-Yuan and Zhang, Wei-Min and Jin, Jinshuang},
  journal = {Phys. Rev. B},
  volume = {83},
  issue = {11},
  pages = {115318},
  numpages = {5},
  year = {2011},
  month = {Mar},
  publisher = {American Physical Society},
  doi = {10.1103/PhysRevB.83.115318},
  url = {https://link.aps.org/doi/10.1103/PhysRevB.83.115318}
}

@article{PhysRevB.86.115453,
  title = {Transient quantum transport in double-dot Aharonov-Bohm interferometers},
  author = {Tu, Matisse Wei-Yuan and Zhang, Wei-Min and Jin, Jinshuang and Entin-Wohlman, O. and Aharony, A.},
  journal = {Phys. Rev. B},
  volume = {86},
  issue = {11},
  pages = {115453},
  numpages = {10},
  year = {2012},
  month = {Sep},
  publisher = {American Physical Society},
  doi = {10.1103/PhysRevB.86.115453},
  url = {https://link.aps.org/doi/10.1103/PhysRevB.86.115453}
}

@article{PhysRevB.86.195403,
  title = {Coherent control of double-dot molecules using Aharonov-Bohm magnetic flux},
  author = {Tu, Matisse Wei-Yuan and Zhang, Wei-Min and Nori, Franco},
  journal = {Phys. Rev. B},
  volume = {86},
  issue = {19},
  pages = {195403},
  numpages = {6},
  year = {2012},
  month = {Nov},
  publisher = {American Physical Society},
  doi = {10.1103/PhysRevB.86.195403},
  url = {https://link.aps.org/doi/10.1103/PhysRevB.86.195403}
}

@article{PhysRevB.85.155324,
  title = {Dynamics of coherences in the interacting double-dot Aharonov-Bohm interferometer: Exact numerical simulations},
  author = {Bedkihal, Salil and Segal, Dvira},
  journal = {Phys. Rev. B},
  volume = {85},
  issue = {15},
  pages = {155324},
  numpages = {10},
  year = {2012},
  month = {Apr},
  publisher = {American Physical Society},
  doi = {10.1103/PhysRevB.85.155324},
  url = {https://link.aps.org/doi/10.1103/PhysRevB.85.155324}
}

@article{Roy,
author = {Roy, Souvik and Maiti, Santanu and Laroze, David},
year = {2025},
month = {06},
pages = {},
title = {Anomalous persistent current in a 1D dimerized ring with aperiodic site potential: Non-interacting and interacting cases},
volume = {96},
journal = {Chinese Journal of Physics},
doi = {10.1016/j.cjph.2025.05.024}
}

@unknown{Mondal,
author = {Mondal, Moumita and Maiti, Santanu},
year = {2025},
month = {04},
pages = {},
title = {Phonon mediated spin polarization in a one-dimensional Aubry-Andre-Harper chain},
doi = {10.48550/arXiv.2504.05906}
}

@article{Roy2023,
  author  = {Roy, Souvik and Ganguly, Sudin and Maiti, Santanu K.},
  title   = {Interplay between hopping dimerization and quasi-periodicity on flux-driven circular current in an incommensurate Su--Schrieffer--Heeger ring},
  journal = {Scientific Reports},
  volume  = {13},
  number  = {1},
  pages   = {4093},
  year    = {2023},
  doi     = {10.1038/s41598-023-31354-9}
}

@article{9yh2-lkjp,
  title = {Flux-driven charge and spin transport in a dimerized Hubbard ring with Fibonacci modulation},
  author = {Roy, Souvik and Padhi, Soumya Ranjan and Mishra, Tapan},
  journal = {Phys. Rev. B},
  volume = {113},
  issue = {15},
  pages = {155401},
  numpages = {11},
  year = {2026},
  month = {Apr},
  publisher = {American Physical Society},
  doi = {10.1103/9yh2-lkjp},
  url = {https://link.aps.org/doi/10.1103/9yh2-lkjp}
}

@article{5vmn-vsxf,
  title = {Non-Hermitian comb effect in coupled clean and quasiperiodic chains},
  author = {Padhi, Soumya Ranjan and Roy, Souvik and Paul, Biswajit and Banerjee, Sanchayan and Mishra, Tapan},
  journal = {Phys. Rev. B},
  pages = {},
  year = {2026},
  month = {Jun},
  publisher = {American Physical Society},
  doi = {10.1103/5vmn-vsxf},
  url = {https://link.aps.org/doi/10.1103/5vmn-vsxf}
}

\end{document}